\documentclass[aip,jmp,amsmath,amssymb,
]{revtex4-1}

\usepackage{float}
\usepackage{amsmath,amssymb,setspace}
\usepackage{subfigure}
\usepackage{chngcntr} %# changes the figure numbering from that point on so that the section number is included and resets the %#numbering of figures at the beginning each subsequent section.
\usepackage{color}
\usepackage{xcolor}
\usepackage{braket}
\usepackage{algorithm}
\usepackage{algpseudocode}
\usepackage{rotating}

\usepackage{comment}

\usepackage{graphicx}

\usepackage{mathabx}

\makeatletter
\def\l@subsection#1#2{}
\def\l@subsubsection#1#2{}
\makeatother
\begin{document}

\title{Scaling Up Electronic Structure Calculations on Quantum Computers: The Frozen Natural Orbital Based Method of Increments}
\date{\today}

\author{Prakash Verma}
\affiliation{1QB Information Technologies (1QBit),  200-1285 W Pender St, Vancouver, BC, V6E 4B1, Canada}

\author{Lee Huntington}
\affiliation{1QB Information Technologies (1QBit),  200-1285 W Pender St, Vancouver, BC, V6E 4B1, Canada}

\author{Marc P. Coons}
\affiliation{Dow, Core R\&D, Chemical Science, 1776 Building, Midland, MI, 48674, USA}

\author{Yukio Kawashima}
\affiliation{1QB Information Technologies (1QBit),  200-1285 W Pender St, Vancouver, BC, V6E 4B1, Canada}

\author{\\Takeshi Yamazaki}
\email{takeshi.yamazaki@1qbit.com}
\affiliation{1QB Information Technologies (1QBit),  200-1285 W Pender St, Vancouver, BC, V6E 4B1, Canada}

\author{Arman Zaribafiyan}
\affiliation{1QB Information Technologies (1QBit),  200-1285 W Pender St, Vancouver, BC, V6E 4B1, Canada}

	\begin{abstract}
The method of increments and frozen natural orbital (MI-FNO) framework is introduced  to help expedite the application of  noisy, intermediate-scale quantum~(NISQ) devices for quantum chemistry simulations. The MI-FNO framework provides a systematic reduction of the occupied and virtual orbital spaces for quantum chemistry simulations. The correlation energies of the resulting increments from the MI-FNO reduction can then be solved by various algorithms, including quantum algorithms such as the phase estimation algorithm and the variational quantum eigensolver (VQE). 
The unitary coupled-cluster singles and doubles VQE framework is used to obtain correlation energies for the case of small molecules (i.e., BeH$_2$, CH$_4$, NH$_3$, H$_2$O, and HF) using the cc-pVDZ basis set. 
The quantum resource requirements are estimated for a constrained geometry complex (CGC) catalyst that is utilized in industrial settings for the polymerization of $\alpha$-olefins. We show that the MI-FNO approach provides a significant reduction in the qubit  requirements relative to the full system simulations.
We propose that the MI-FNO framework can create scalable examples of quantum chemistry problems that are appropriate for assessing the progress of NISQ devices.
	\end{abstract}

\maketitle

\section{Introduction}
Accurate characterization of electron interactions is vital for the computational design of molecules and requires finding exact solutions of the electronic Schr\"odinger equation.
Solving  Schr\"odinger's equation  exactly on  classical computers is a computationally demanding task because the dimension of the Hilbert space of quantum systems increases exponentially with system size and the complexity of finding exact solutions scales factorially with the number of orbitals and electrons.
Thus, on classical hardware, obtaining solutions of Schr\"odinger's equation is only possible for small systems.\cite{Head-Gordon-2008-58} 

In recent years, there has been increasing interest in quantum computation, a new computing paradigm initially conjectured as an efficient framework for simulating quantum mechanical systems.\cite{Yuri-1980,Feynman1982} 
In the decades since this conjecture was put forward, there has been tremendous theoretical progress towards realizing the concept of using a quantum computer for quantum simulations.\cite{Lloyd1996}
Early implementations of quantum algorithms aimed at applications in computational chemistry\cite{Aspuru-Guzik2005} were deployed on quantum computers in order to evaluate molecular energies.\cite{PeruzzoPhotonicPQ2014,HMR2018, OBK2016, KMT2017, IonQWater2017}
There has also been accelerated progress in hardware development.
For example, IBM,\cite{IBM} Google,\cite{Google} Intel,\cite{Intel} Rigetti, \cite{Rigetti} and QCI\cite{QCI} have all developed quantum computing platforms based on superconducting qubits, while IonQ \cite{IonQ} and Honeywell\cite{Honeywell} have developed platforms based on ion traps. Google’s achievement on a benchmarking milestone commonly referred to as ``quantum supremacy''~\cite{GoogleQuantumSupremacy2019} demonstrates a transition of the quantum computing field away from a purely theoretical concept.
Despite the progress of hardware development, current quantum devices are error-prone and have limited computing capacity, hence the introduction of the term noisy, intermediate-scale quantum (NISQ)\cite{NISQ_Preskill2018} to describe them.

The limitations of NISQ hardware have driven significant progress towards the development of algorithms that seek to shorten the timescale for the successful application of quantum computers for solving quantum chemistry problems.
Some of these developments include quantum--classical hybrid algorithms for variational optimization\cite{Peruzzo:2014aa,Ganzhorn-2018-1809.05057,Nakanishi-2018-1810.09434,Matsuura-2018-1810.11511} and wave function ans{\"{a}}tze that produce low-depth circuits for efficient quantum simulation.\cite{OMalley:2016aa,Romero-2017-1701.02691,KMT2017,Babbush-2017-1706.00023,Kivlichan-2017-1711.04789,Grimsley-2018-1812.11173,Ryabinkin:2018ab,Ryabinkin-2019-1906.11192}
Research has also focused on incorporating problem decomposition (PD) techniques developed for applications of classical quantum chemistry~\cite{Bauer-2016-1510.03859,Rubin-2016-1610.06910,KMT2017,Hempel-2018-1803.10238,Kuhn-2018-1812.06814,Yamazaki-2018-1806.01305,Gao-2019-1906.10675,mochizuki_okuwaki_kato_minato_2019,Gonthier:2020} into quantum algorithms to further improve the efficiency of simulations on NISQ devices.
The advantage of PD techniques is the ability to decompose the full electronic structure problem of a molecule into a set of smaller sub-problems that can be solved more efficiently. Problem decomposition approaches also provide a good approximation to the results of calculations performed on the corresponding full system.
Such approaches have a long history in the literature, originating from early investigations of local electron correlation by Sinano{\u{g}}lu,\cite{Sinanoglu} Nesbet,\cite{Nesbet_local} and Ahlrichs and Kutzelnigg\cite{local_ahlrichs} during the 1960s.
Several comprehensive reviews of PD techniques used in quantum chemistry applications are available in Refs.~\citenum{fragment_based_methods,fragment_based_methods_2,Sun:2016aa,embedding_methods}.

Problem decomposition techniques reduce the effective problem size of a molecular system and create opportunities to characterize near-term devices by enabling hardware experiments for larger systems that would otherwise be inaccessible during the NISQ era.
Following a similar strategy as some of the authors' previous work,~\cite{Yamazaki-2018-1806.01305} we seek an efficient methodology for performing scalable quantum chemistry simulations on near-term devices based on PD techniques.
Specifically, we explore a strategy that combines the method of increments (MI)~\cite{PhysRev.155.51,PhysRev.155.56,PhysRev.175.2} with the frozen natural orbital (FNO) approach\cite{FNO} to achieve an MI-FNO framework that enables a two-fold reduction of the occupied and virtual molecular orbital (MO) spaces, respectively.
The systematic truncation of the occupied MO space also becomes essential, in particular, when larger molecular systems that have a considerable number of electrons are targeted.
We achieve a reduction in the occupied MO space by adapting a recently proposed incremental full configuration interaction (iFCI) approach~\cite{Zimmerman:2017ab,Zimmerman:2017ac,Zimmerman:2017aa} that is based on the method of increments and provides a polynomial scaling approximation to full configuration interaction (FCI).
By decomposing the problem into \mbox{$n$-body} sub-problems (or ``increments''), it has been shown that accurate correlation energies can be recovered at low values of $n$ in a highly parallelizable computation.~\cite{Zimmerman:2017ab,Zimmerman:2017ac,Zimmerman:2017aa}
The success of other incremental approaches has also been demonstrated elsewhere for traditional quantum chemistry applications.~\cite{Stoll:1992aa, Modll:1997aa, Bezugly:2004aa, Stoll:2005aa, Friedrich:2007aa, Dahlke2007aa, Bytautus:2010aa, Gordon:2012ab, Mueller:2012aa, Richard:2012, Friedrich:2013a, Friedrich:2013b, Zhang:2013ab, Voloshina:2014aa, Anacker:2016, Lao:2016aa, Fiedler:2017, Eriksen:2017aa, Boschen:2017aa, Eriksen:2018, Fertitta:2018aa, Zimmerman:2019aa, Eriksen:2019a, Eriksen:2019b}
Furthermore, several recent investigations~\cite{Bauer-2016-1510.03859,Rubin-2016-1610.06910,Reiher:2017aa,KMT2017,Yamazaki-2018-1806.01305,Hempel-2018-1803.10238,Kuhn-2018-1812.06814,Gao-2019-1906.10675,mochizuki_okuwaki_kato_minato_2019,Gonthier:2020} have focused on reducing the complexity of chemical systems on quantum computers by utilizing active spaces (i.e., ignoring certain occupied and virtual space orbitals) or truncating the virtual orbital space (i.e., removing higher eigenvalue canonical virtual orbitals or systematically reducing  the virtual space based on the FNO approach).

A similar strategy to ours is deployed by Fielder et al.,~\cite{Fiedler:2017} where an incremental scheme is combined with local-pair natural orbitals~\cite{Neese_1,Neese_2,Neese_3,Neese_4,Neese_5,Hattig_PNO,Werner_PNO} to achieve highly efficient and accurate reaction energies for large molecular systems.
However, to our knowledge, an approach for systematically and simultaneously reducing the occupied and virtual MO spaces has not yet been utilized for applications of quantum chemistry on NISQ devices.
Here, we demonstrate how the method of increments can be used to reduce the occupied space of a molecular system, and FNOs can be employed to truncate the virtual space.
As a first step, we validate the accuracy of the MI-FNO approach, and demonstrate its ability to reduce both the occupied and virtual spaces while maintaining a reasonable level of accuracy, by examining the small molecules BeH$_2$, CH$_4$, NH$_3$, H$_2$O, and HF using a moderate-sized cc-pVDZ basis set.~\cite{Dunning:1989aa}
The ability to solve the electron correlation problem on classical (conventional) computers depends on the size of the computational space of the molecule.
The molecular computational space can be defined in terms of the number of electrons or occupied orbitals and the total number of MOs. For quantum devices, the corresponding computational space can be represented in terms of the number of qubits, and the number of one- and two-qubit gates. In order to map the electronic structure problem onto a quantum device, the Fock space representation of the wavefunction is used and the wavefunction is evolved using quantum gates. The number of qubits is equal to the number of molecular spin orbitals, while the complexity of the wave function can be represented by the number of  one- and two-qubit gates. 
To demonstrate the efficacy of our MI-FNO approach on larger molecules,  we provide a qubit count estimation for an industrially relevant $\alpha$-olefin polymerization catalyst, a constrained geometry complex (CGC) catalyst,~\cite{Arriola:2007aa} using the cc-pVDZ and cc-pVTZ basis sets.~\cite{Dunning:1989aa}

This paper is organized as follows. In Sec.~\ref{sec:theory}, a review of the MI-FNO approach within a variational quantum eigensolver (VQE) framework is provided as an example of the quantum solvers suitable for NISQ devices.
In Sec.~\ref{sec:detail}, the computational details are described and a schematic illustration is provided of our MI-FNO approach for large-scale quantum chemistry simulations on quantum hardware.
In Sec.~\ref{sec:r_and_d}, we present the resulting molecular energies obtained using the MI-FNO approach and discuss its applicability for use on near-term devices.
Sec.~\ref{sec:conclusion}  provides a summary of results and possibilities for future work.

\section{Theory} \label{sec:theory}
\begin{figure}[h]
	\includegraphics[scale=0.45]{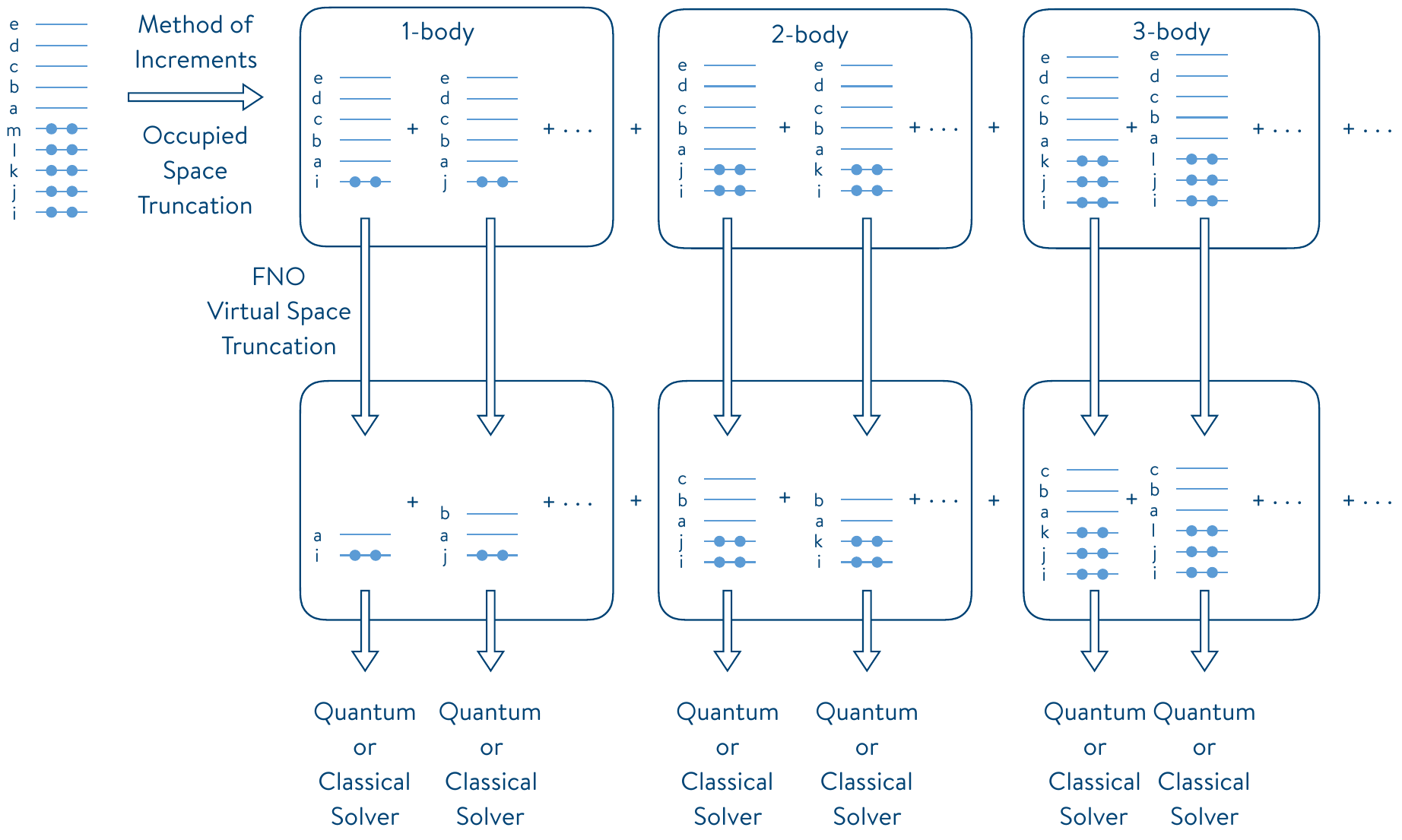}
	\caption{Conceptual  and  schematic  illustration  of  the MI-FNO framework for scaling up the size of molecules for quantum chemistry simulations on quantum hardware. The full molecular computational space is defined in terms of occupied orbitals (i, j, k,...) and virtual orbitals (a, b, c,...). The method of increments decomposes the occupied space into smaller 1-, 2-, and 3-body increments which consists of one, two, and three occupied orbitals, respectively. The virtual space of a given $n$-body increment can be reduced by using the FNO procedure.}
	\label{fig:MI-FNO}
\end{figure}

The MI-FNO approach provides a framework for dividing the occupied  space of a molecule using MI while the corresponding virtual space of each increment is compacted separately using the FNO procedure. In this section, we provide a brief overview of the ingredients that make up the framework, namely, MI to decompose the full occupied space into 1-, 2-, and 3-body increments (see Figure \ref{fig:MI-FNO})
and the use of FNOs to design a tailored  virtual space for each increment. The electron correlation problem can then be solved in a reduced computational space using a given quantum algorithm, or a conventional quantum chemistry approach when needed. The VQE,  coupled with the unitary coupled-cluster singles and doubles (UCCSD) wavefunction ansatz, is explored as an example of a possible quantum approach for NISQ devices.

\subsection{Conventional Problem Decomposition Techniques}
\subsubsection{Reducing the Occupied Space: The Method of Increments}
The MI approach relies on the many-body or $n$-body Bethe--Goldstone expansion~\cite{Bethe-Goldstone}  of the electron correlation energy of a molecule. It was first introduced in quantum chemistry by Nesbet.~\cite{PhysRev.155.51,PhysRev.155.56,PhysRev.175.2}  The electron correlation energy ($E_\text{c}$) is defined as the difference between the exact ($E_\text{exact}$) and the Hartree--Fock (mean-field) energy ($E_\text{HF}$).  Using the many-body expansion, the electron correlation energy  can be expressed in terms of $n$-body increments ($\epsilon_{i} $, $\epsilon_{ij}$, $\epsilon_{ijk}$, and $\epsilon_{ijkl}$) as

\begin{align}\label{Ec_dec}
	E_\text{c} &= E_{\text{exact}} - E_{\text{HF}}\nonumber\\
	&=\sum_i \epsilon_{\text{i}} + \sum_{i>j} \epsilon_{ij} + \sum_{i>j>k} \epsilon_{ijk} +\sum_{i>j>k>l} \epsilon_{ijkl} \ldots
\end{align}

The $n$-body increments  $\epsilon_{i} $, $\epsilon_{ij}$, $\epsilon_{ijk}$, and $\epsilon_{ijkl}$ are, respectively, the one-, two-, three-, and four-body increments defined as

\begin{align}
	\epsilon_{i} &= E_\text{c}(i)\\
	\epsilon_{ij} &= E_\text{c}(ij) - \epsilon_{i} - \epsilon_{j}\\
	\epsilon_{ijk} &= E_\text{c}(ijk) - \epsilon_{ij} - \epsilon_{ik} - \epsilon_{jk} - \epsilon_{i} - \epsilon_{j} - \epsilon_{k}\\
	\epsilon_{ijkl} &= E_\text{c}(ijkl) - \epsilon_{ijk} - \epsilon_{ijl} - \epsilon_{jkl} - \cdots\\
	&\vdots\nonumber
\end{align}

\noindent where $E_\text{c}(i)$ denotes the correlation energy of the increment $i$, $E_\text{c}(ij)$ denotes the correlation energy of the increment $i,j$, and so on.
These increments (the indices (\(i\), \(j\), \(k\), \(\ldots\))) appearing in the expansion of Eq.~(\ref{Ec_dec}) can be orbitals, atoms, molecules, or fragments.\cite{Stoll:1992aa, Modll:1997aa, Bezugly:2004aa, Stoll:2005aa, Friedrich:2007aa, Dahlke2007aa, Bytautus:2010aa, Gordon:2012ab, Mueller:2012aa, Richard:2012, Zhang:2013ab, Voloshina:2014aa, Anacker:2016, Lao:2016aa, Zimmerman:2017ab, Eriksen:2017aa, Boschen:2017aa, Eriksen:2018, Fertitta:2018aa, Zimmerman:2019aa, Eriksen:2019a, Eriksen:2019b}

The 1-body increments include one doubly occupied orbital, 2-body increments include two distinct doubly occupied orbitals, and so on.
Depending on the nature of the correlation problem and/or the available computational resources, any suitable algorithm can be chosen to predict the correlation energies, whether geared towards quantum computing or classical architectures. The present framework can work with several quantum algorithms, such as VQE~\cite{Peruzzo:2014aa}  and phase estimation algorithm (PEA).~\cite{Abrams:1997aa, Aspuru-Guzik:2005aa} Note that the MI framework takes advantage of the CC~\cite{Friedrich:2007aa,Zhang:2013ab,Anacker:2016} or FCI approaches.~\cite{Eriksen:2017aa, Zimmerman:2017ab,Zimmerman:2017aa,Zimmerman:2017ac,Eriksen:2018, Eriksen:2019a, Eriksen:2019b}

\subsubsection{Reducing the Virtual Space: Frozen Natural Orbitals}
The method of increments is a technique that provides an efficient and accurate approach for computing electronic correlation energies.
One can view the MI approach as a framework for reducing the full occupied space of a molecule into a much smaller space. However, further reduction of the computational space will be required when we target applications on NISQ computers. In the present study, we incorporate the FNO approach~\cite{FNO_taube_lamdaCCSD,FNO_taube1,Taube:2008ab,FNO_bartlett,doi:10.1063/1.453884} into our framework to further reduce the problem size by truncating the virtual orbital space. In recent work, this approach has been applied to reduce the computational cost of quantum chemistry calculations in quantum computing.~\cite{mochizuki_okuwaki_kato_minato_2019}

The natural orbitals are obtained by diagonalizing the  one-particle reduced density matrix (RDM). 
In the case of ``frozen'' natural orbitals,\cite{FNO_Lowdin} only the virtual--virtual block of the one-particle RDM is diagonalized. The occupied orbitals are just the original canonical Hartree--Fock occupied orbitals while a new transformation and ranking is obtained for the virtual space, such that the reference energy is invariant to the transformation. The correlation energy is also invariant to the transformation if the method employed is invariant under rotations of the virtual orbitals. 
Frozen natural orbitals are considered transformed and ranked virtual MOs that can be obtained at any arbitrary level of an \textit{ab initio} theory. In this work,  we use the MBPT(1) wavefunction, which constitutes the first-order correction to the Hartree--Fock wavefunction. The one-particle virtual--virtual block of the MBPT(2) density matrix is diagonalized to obtain natural orbitals as eigenvectors and corresponding occupation numbers as eigenvalues. These eigenvalues can be used to truncate the virtual space, while the eigenvectors are employed to transform the virtual space. By choosing a certain threshold or population percentage criterion, a certain number of virtual orbitals can be kept, and the rest are ignored. 

The correlation energy is calculated only in the truncated virtual space, and then, the correction term $\Delta E^{\text{MBPT(2)}} = E^{\text{MO}}_{\text{MBPT(2)}}-E^{\text{FNO}}_{\text{MBPT(2)}}$ is added to the correlation energy to recover the full correlation energy.  The correction term $\Delta E^{\text{MBPT(2)}}$ is the MBPT(2) correlation energy in the full molecular orbital space minus the MBPT(2) correlation energy in the truncated FNO space.

In the spin-orbital basis, the virtual--virtual  ($D_{ab}$) block of the one-particle MBPT(2) density matrix~\cite{Taube:2008ab} is expressed as

\begin{equation}\label{key}
	D^{(2)}_{ab}=\frac{1}{2}\sum_{cij}\frac{\langle cb||ij \rangle \langle ij || ca \rangle}{\epsilon^{cb}_{ij} \epsilon^{ca}_{ij}}\,,
\end{equation}
where the quantity $\epsilon^{ab}_{ij}$ in the denominator is defined as $\epsilon^{ab}_{ij} = f_{ii}+f_{jj}-f_{aa}-f_{bb}$, in which \(f\) is the Fock matrix. Note that $\langle cb||ij \rangle = \langle cb | ij \rangle - \langle cb | ji \rangle$ is an antisymmetric two-electron integral. The indices $i$, $j$ represent occupied spin orbitals, while $a$, $b$, and $c$ represent virtual spin orbitals.

\subsection{A Quantum Approach to Electron Correlation}
\subsubsection{The Variational Quantum Eigensolver Algorithm}
We consider the VQE algorithm \cite{Peruzzo:2014aa} as an example of the quantum solvers suitable for near-term applications on NISQ devices. The VQE algorithm was originally introduced, within the context of quantum chemistry, as a hybrid quantum--classical algorithm for solving the molecular electronic Schr{\"o}dinger equation.  According to the variational principle, for a (normalized) parametrization of the wavefunction \(\vert \Psi(\vec{\theta})\rangle\), if one minimizes the expectation value of the Hamiltonian operator \(\widehat{H}\),

\begin{equation}
	E = \langle \widehat{H} \rangle = \min_{\vec{\theta}}\langle \Psi(\vec{\theta})\vert 
	\widehat{H} \vert \Psi(\vec{\theta})\rangle \geq E_{\text{exact}}\,,
	\label{eq:vqe}
\end{equation}
an upper bound to the exact ground state energy is obtained.

We wish to estimate values of the parameters \(\{\theta_1, \theta_2, \ldots, \theta_p\}\) (i.e., the elements of the vector \(\vec{\theta}\)) that minimize the expectation value according to Eq. \ref{eq:vqe}. The VQE algorithm requires a Hamiltonian operator in qubit form (i.e., written in terms of Pauli operators). Furthermore, a unitary parametric ansatz for the wavefunction in the qubit basis is required for the state preparation. Once the initial state has been prepared (i.e., an appropriate set of initial parameters has been used), an expectation value measurement is performed using quantum hardware or an appropriate simulation tool. Subsequently, the current value of the expectation value is fed to a classical optimizer in order to estimate a new set of variational parameters. This  provides a new wavefunction, and the procedure is repeated until an optimized wavefunction and expectation value have been obtained.

The VQE algorithm constitutes a reduced circuit-depth hybrid quantum--classical methodology for solving the molecular electronic Schr{\"o}dinger equation, as it minimizes the use of quantum hardware resources. In the second-quantization picture, the molecular electronic Hamiltonian takes the form

\begin{align}
	\widehat{H} = \sum_{p,q}h_q^p \hat{a}^{\dag}_p a_q + \frac{1}{2} \sum_{p,q,r,s} h_{rs}^{pq} \hat{a}^{\dag}_p\hat{a}^{\dag}_q\hat{a}_s\hat{a}_r\,,\label{MB_hamiltonian}
\end{align}

\noindent in which \(p\), \(q\), \(r\), and \(s\) label general spin-orbitals, and \(a^{\dag}_p\) and \(a_p\) are, respectively, creation and annihilation operators associated with orbital \(p\).  The one- and two-electron integrals, \(h^{p}_{q}\) and \(h^{pq}_{rs}\), are

\begin{align}
	h^{p}_{q}&=\left\langle p\right\vert\widehat{h}\left\vert q \right\rangle=\int\varphi^{*}_p(\mathbf{x})\left( -\frac{1}{2}\nabla^2 -\sum_{\mu=1}^{\mathcal{N}}\frac{Z_{\mu}}{\left\vert
		\mathbf{r}-\mathbf{R}_{\mu}\right\vert } \right)\varphi_q(\mathbf{x})\,\mathrm{d}\mathbf{x}
\end{align}
and
\begin{align}
	h^{pq}_{rs}&=\left\langle pq|rs\right\rangle =\int\varphi^{*}_p(\mathbf{x}_1)\varphi^{*}_q(\mathbf{x}_2)\frac{1}{r_{12}}\varphi_r(\mathbf{x}_1)\varphi_s(\mathbf{x}_2)\,\mathrm{d}\mathbf{x}_1\mathrm{d}\mathbf{x}_2\,,
\end{align}

\noindent respectively, in which \(Z_{\mu}\) and \(R_{\mu}\) are the charge and position of nucleus \(\mu\), respectively, and \(r_{12} = \vert \mathbf{r}_2 - \mathbf{r}_1 \vert\) is the inter-electronic distance. The molecular Hamiltonian can be transformed into the qubit basis by using the Jordan--Wigner transformation,~\cite{Jordan:1928aa}

\begin{equation}
	\widehat{H} = \sum_p h_p^{\alpha} \sigma_p^{\alpha} + \sum_{pq}h_{pq}\sigma_{p}^{\alpha}\sigma_{q}^{\beta} + \sum_{pqr}h_{pqr}\sigma_{p}^{\alpha}\sigma_{q}^{\beta}\sigma_{r}^{\gamma} + \ldots, \label{qubit_H}
\end{equation}

\noindent or another available transformation technique (e.g., Bravyi--Kitaev,~\cite{Bravyi:2002aa} Bravyi--Kitaev
\newline Superfast~\cite{Setia:2018aa}). Here, \(p\), \(q\), \(r\),\(\dots\) label qubits, and \(\sigma_p^{\alpha}\), where \(\alpha \in x,y,z\) is a Pauli matrix acting on qubit~\(p\).

\subsubsection{The Unitary Coupled-Cluster Ansatz}
While there are several strategies for deriving a parametric ansatz for the wavefunction (e.g., hardware efficient,\cite{KMT2017} QCC,\cite{Ryabinkin:2018ab} and iQCC\cite{Ryabinkin-2019-1906.11192}), we  consider the UCC ansatz in this work. The choice of ansatz is  important for the convergence of the classical optimization and has a marked effect on the circuit depth. The latter issue is beyond the scope of the present study, but we plan to return to it in future work. Let us assume that the Hartree--Fock equations have been solved to obtain a zeroth-order, single-determinantal, mean-field wavefunction \(\vert\Psi_0\rangle\) and the one- and two-electron integrals in the spin-orbital basis. The UCC ansatz for the correlated wavefunction can then be written as

\begin{equation}
	\Psi({\vec{\theta}}) = e^{\widehat{T} - \widehat{T}^{\dag}} \vert \Psi_{0}\rangle \label{ucc_ansatz}\,,
\end{equation}
in which the cluster operator is defined as
\begin{align}
	\widehat{T} &= \widehat{T}_1 + \widehat{T}_2 + \ldots\\
	&= \sum_{i,a}\theta_i^a \hat{a}^{\dag}_a\hat{a}_i + \frac{1}{2} \sum_{i,j,a,b}\theta_{ij}^{ab}\hat{a}^{\dag}_a\hat{a}^{\dag}_b\hat{a}_j\hat{a}_i + \ldots \label{cluster_op}
\end{align}

The UCC ansatz is usually truncated up to double excitations (i.e., including only \(\widehat{T}_1\) and \(\widehat{T}_2\) in Eq.~\ref{cluster_op}), thus defining UCCSD. In analogy with the Hamiltonian in Eq.~\ref{qubit_H}, the ansatz of Eq.~\ref{ucc_ansatz} can be transformed into the qubit basis. Due to the non-commuting nature of the operators used in the UCCSD ansatz, the Suzuki--Trotter decomposition is used to decompose the exponential of the cluster operator as a product of unitary operators acting on the reference wavefunction (obtained from a classical Hartree--Fock calculation), and is  subsequently transformed into a qubit representation. This Trotterized UCCSD ansatz is then used for the state preparation step of the VQE algorithm discussed above to find an approximate expectation value of the molecular electronic Hamiltonian, thus providing an estimate of the ground-state energy of a given molecule.

\section{Computational details} \label{sec:detail}
We perform UCCSD calculations using the incremental expansion approach and FNO-based virtual space truncation (MI-FNO-UCCSD). In order to understand the convergence behaviour of MI-FNO-UCCSD energies as the size of the computational space grows, we also perform the MI-FNO calculation using  conventional CCSD. The calculations are performed on the experimental molecular geometries of BeH\(_2\), CH\(_4\), NH\(_3\), H\(_2\)O, and HF obtained from the NIST Computational Chemistry Comparison and Benchmark Database.~\cite{NIST} The cc-pVDZ basis set~\cite{Dunning:1989aa} is used for all of the calculations.

In the incremental expansion approach, we consider the many-body expansion series including up to two-body terms for BeH\(_2\), as the expansion including up to three-body terms is equivalent to solving the full problem (i.e., BeH\(_2\) has three doubly occupied orbitals). The expansion up to two-body terms for BeH\(_2\) includes three one-body increments and three two-body increments---in total, six increments. For the rest of the molecules, which have five occupied orbitals, we examine the expansions up to three- and four-body increments. The resulting total numbers of increments are 25 and 30, respectively, for the expansions. For the virtual orbitals of each increment, we examine the effect of the size of the virtual space by adding one virtual orbital (which has a higher FNO occupancy) at a time, to the computational space of the increments. The implementation of the \mbox{MI(\(n\))-FNO} approach is numerically validated by comparing the total energies computed with the MI(\(n\))-FNO-CCSD approach with full system CCSD energies. The results are provided in Appendix~\ref{appendix:appA}.

For each molecular system except BeH$_2$, the electronic structure problems for each of the increments, with a truncated virtual space up to five virtual orbitals, is solved by using VQE with the UCCSD ansatz, leading to, at most, a 16-qubit problem. For BeH$_2$,  the expansion including two-body terms along with seven virtual orbitals is considered, which leads to, at most, an 18-qubit problem.
An MBPT(2) FNO correction is added to the correlation energies obtained using a truncated virtual space, in order to account for the missing correlation energies. The resulting correlation energies for each increment are used to reconstruct the correlation energy of the entire molecule by following the expansion scheme described in the previous section. 
We refer to the present approach as \mbox{MI(\(n\))-FNO-UCCSD} (or \mbox{MI(\(n\))-FNO-CCSD} if the classical CCSD approach is used to obtain the correlation energy), where \(n\) indicates the expansion up to \(n\)-body increments. 
To obtain an estimate of the number of qubits for a molecule relevant to industry, 
we consider a CGC catalyst utilized in the polymerization of $\alpha$-olefins.~\cite{Arriola:2007aa} The configuration of the catalyst is obtained from a crystal structure of the CGC catalyst and the cc-pVDZ and the cc-pVTZ basis sets are utilized.~\cite{Dunning:1989aa}

All of the quantum simulations reported are performed using the OpenFermion\cite{OpenFermion} and ProjectQ\cite{Steiger2018projectqopensource} software packages and the OpenFermion-ProjectQ\cite{OpenFermion} interface. 
The molecular integrals and Hartree--Fock solutions are generated using PySCF.\cite{Sun:2018aa} Incremental decomposition of the occupied orbitals and the corresponding generation of scalable FNO-transformed virtual space is achieved using the development version of \textit{QEMIST Cloud}, the \textit{Quantum-Enabled Molecular Ab Initio Simulation Toolkit}.\cite{QEMIST}
The OpenFermion program package is employed to map second-quantized quantities (e.g., Hamiltonian, UCCSD ansatz) to the qubit basis. 
The qubit representation of the truncated molecular Hamiltonian is obtained using the Jordan--Wigner transformation\cite{Jordan:1928aa} implemented in OpenFermion. The VQE simulations, using the UCCSD ansatz, are performed using ProjectQ and OpenFermion-ProjectQ. 
The OpenFermion-ProjectQ interface is then employed to convert the qubit form of the UCCSD ansatz into a Trotterized time evolution operator, which can easily be expressed in terms of elementary universal quantum gate operations. 
The ProjectQ program package, which is an ideal (noiseless) state vector simulator, is used to simulate the UCCSD circuits and to evaluate the expectation value of the qubit Hamiltonian in the UCCSD state (i.e., using the exact representation of the state vector). ProjectQ is also employed to perform the gate counts using its resource estimation utility. 
The classical optimization steps of VQE are performed using the COBYLA algorithm\cite{Powell:1994aa} with a convergence tolerance of $10^{-5}$. The MBPT(1) amplitudes are used as an initial guess of the parameters for the UCCSD trial wavefunction. The conventional CCSD energy of the full problem is also calculated using PySCF and used as a reference energy. In performing the conventional CCSD calculation, we use a tolerance of $10^{-7}$ hartrees.

\section{Results and Discussion} \label{sec:r_and_d}
\subsection{The Quantum Computational Efficiency of the MI-FNO Approach}
Noisy, intermediate-scale quantum  hardware is limited not only in the number of qubits it has, but also in the number of gate operations. Therefore, it is important to understand the amount of quantum resources  needed to achieve the desired accuracy in  electronic structure calculations.~\cite{Kuhn-2018-1812.06814} In this section, we discuss to what extent the MI(\(n\))-FNO approach can reduce the quantum resources compared to full UCCSD simulation. The number of one- and two-qubit gates we report should be considered as an upper bound of the gate counts, as the actual number of gate counts can vary depending on the level of circuit optimization.

The MI-FNO approach provides a framework for decomposing complex quantum systems into smaller increments that can easily be simulated or computed on NISQ devices. One can apply not only the UCC wavefunction on quantum devices but also the conventional coupled-cluster approach on classical machines to obtain electron correlation energies in the truncated computational space. As discussed below, the energy profiles of MI-FNO-UCCSD and MI-FNO-CCSD closely follow each other. This is quite encouraging, as we can now estimate  the accuracy of the UCCSD approach for systems requiring an increasing number of qubits by using the accuracy of the CCSD method. Our goal is to approximate the CCSD energy of the full system by using the MI-FNO-CCSD or MI-FNO-UCCSD approach. We are interested in knowing  the size of virtual space needed to approximate the CCSD energy in the full computational space to within 1~kcal/mol accuracy. The idea is not to provide a magic number for virtual orbitals that will be needed to obtain chemical accuracy, but simply to explore  the possibility of designing scalable examples of quantum chemistry problems that are appropriate for measuring the progress of NISQ devices. A further goal is to answer the question of to what extent quantum computational resources can be reduced with such an aggressive truncation of the virtual space. 

\begin{table} [h]
	\caption {Quantum resources required to obtain chemically accurate energies. The number before the slash represents the quantum resources needed when using the MI(\(n\))-FNO approach, and the number after the slash represents the quantum resources required for full UCCSD simulation without PD. The percentages given in parentheses represent the extent of reduction that the \mbox{MI(\(n\))-FNO} approach achieved.} \label{tab:quantum_resources}
	\renewcommand{\arraystretch}{1.2}
	\setlength{\tabcolsep}{1em}
	\begin{tabular}{ c | c c c}
		& \# of qubits & \# of one-qubit gates & \# of two-qubit gates \\
		\hline
		BeH$_\text{2}$ & 18/48 (63\%) & 4180/73,230 (94\%) &  6944/302,160 (98\%) \\
		CH$_\text{4}$ & 32/68 (53\%) &  143,214/1,726,498 (92\%) &  384,592/9,482,448 (96\%) \\
		NH$_\text{3}$ & 32/58 (45\%) &  103,830/731,602 (86\%)&  275,520/3,373,264 (92\%)\\
		H$_\text{2}$O & 34/48 (29\%) &  62,934/241,138 (74\%)&  182,592/929,648 (80\%)\\
		HF & 26/38 (32\%) &  36,870/205,498 (82\%) & 83,328/660,032 (87\%)\\
	\end{tabular}
\end{table}

Table~\ref{tab:quantum_resources} gives a summary of the quantum resources required to achieve chemically accurate energies. The convergence behaviour of the energies of MI-FNO-UCCSD and MI-FNO-CCSD  with respect to CCSD with an incremental increase of FNO transformed virtual orbitals can be found in  Figures~\ref{fig:Energies_BeH2} and~\ref{fig:Energies_}.  A more detailed discussion can be found in the next section. 
The number of qubits required for the MI-FNO approach to achieve chemical accuracy with respect to the parent CCSD values are 32, 32, 34, and 26 for CH$_\text{4}$, NH$_\text{3}$, H$_\text{2}$O, and HF, respectively, which is a significant reduction, as the number of qubits necessary to perform a direct simulation of the full system are 68, 58, 48, and 38 for CH$_\text{4}$, NH$_\text{3}$, H$_\text{2}$O, and HF, respectively. 

A more detailed breakdown of the quantum resource estimations for MI(\(2\))-FNO-UCCSD with an incremental increase of the virtual space for BeH$_\text{2}$ is shown in Appendix~\ref{appendix:appB} in Figure \ref{fig:Quantum_Resources_BeH2}. While Figures \ref{fig:Quantum_Resources_HF}, \ref{fig:Quantum_Resources_H2O}, \ref{fig:Quantum_Resources_NH3}, and
\ref{fig:Quantum_Resources_CH4} illustrate the resources required to perform \mbox{MI(\(3\))-FNO-UCCSD} calculations on the HF, H$_\text{2}$O, NH$_\text{3}$, and CH$_\text{4}$ molecules, respectively. As the main focus of this study is to design a framework that reduces qubit counts in a  manner such that systematically improvable results can be achieved with increasing quantum computational resources, detailed discussion on the reduction of other quantum resources such as gate counts and measurement we leave for future work. In brief,
we find that our MI(\(n\))-FNO-UCCSD approach considerably reduces the number of gate operation by 74\% to 98\% from those required to perform full UCCSD simulations without PD. 
The resulting number of gate operations remains very large for NISQ hardware; however, our MI(\(n\))-FNO approach is general, and can be combined with any other ansatzes that may provide shallower circuits than UCCSD, such as the ``hardware-efficient'' ansatz~\cite{KMT2017} and QCC methods.~\cite{Ryabinkin:2018ab, Ryabinkin-2019-1906.11192,lang2020unitary}

Due to the larger quantum computational requirements, performing VQE calculations based on the full molecular space UCCSD ansatz are nearly impossible on an existing quantum device. For smaller molecules like BeH$_\text{2}$, HF, H$_\text{2}$O, NH$_\text{3}$, and CH$_\text{4}$, 38 to 68 qubits are needed for the cc-pVDZ basis. Detailed information regarding the number of qubits, one-qubit gates, and  two-qubit gates for these molecules using the cc-pVDZ basis can be found in Table~\ref{tab:quantum_computational_space} in the Appendix~\ref{appendix:appB}.

By limiting many-body expansion of electron correlation energies to three-body increments one can reduce any occupied space to  a maximum of three doubly occupied orbitals.
Assuming the full virtual space is used to perform MI(\(3\))-UCCSD calculations, the qubit requirements for HF, H$_\text{2}$O, NH$_\text{3}$, and CH$_\text{4}$ are reduced by four qubits. Without employing any aggressive strategies for the virtual space truncation, demonstrating the applicability of NISQ devices for chemistry applications is difficult. One can use an active space approach, in which a selected number of Hartree--Fock virtual orbitals are included in the computational space to perform  correlation energy calculations.  Alternatively, one can also rank the virtual orbitals, using FNO occupancies and select the most important virtual orbitals to obtain much improved correlation-energy calculation results in a truncated FNO space (i.e., compared to a truncated canonical Hartree--Fock MO space). It is well-known that wavefunction-based approaches require larger basis sets to effectively capture electron correlation  and the corresponding virtual space of the larger basis is often sparse. Frozen natural orbitals can be used as a tool to recognize the sparsity and to help compress the virtual space. Often up to a \mbox{50--60\%} reduction in the virtual space is possible with a loss of only 1\% in the correlation energy.\cite{FNO_taube1,Taube:2008ab,Landau:2010aa} The effectiveness of the FNO approach in compressing the virtual space can be demonstrated by plotting the cumulative FNO occupancy percentage  as a function of the number of virtual orbitals. A more detailed discussion is presented in the Appendix~\ref{appendix:appA} along with the assembled plots (see Figure~\ref{fig:FNO}). 

\subsection{The Accuracy of the MI(\(n\))-FNO-UCCSD Approach }
To investigate the accuracy of MI($n$)-FNO-UCCSD energies, the  energies of BeH\(_\text{2}\), CH\(_\text{4}\), NH\(_\text{3}\), H\(_\text{2}\)O, and HF are obtained using MI-FNO-UCCSD with  various  truncated virtual spaces.  Although the number of virtual orbitals for each increment is the same, it is worth noting that unique FNO transformations are generated for each increment.

\begin{figure*}[b]
	\includegraphics[scale=.6]{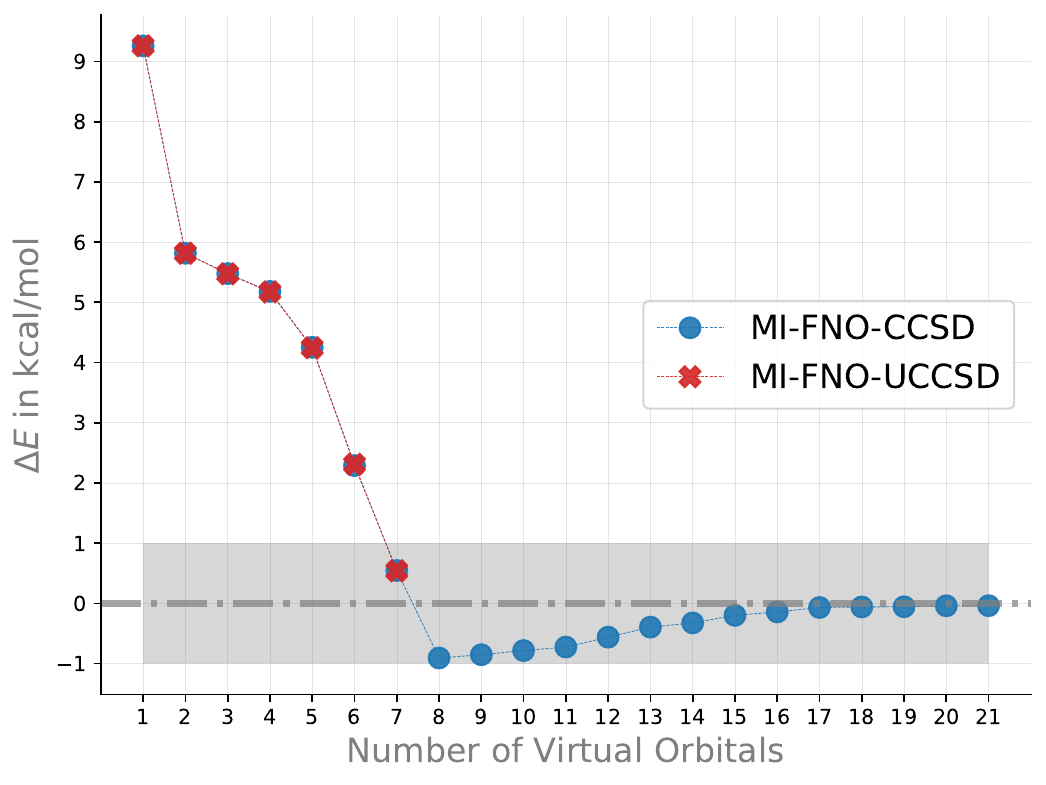}
	\caption{Energy deviation ($\Delta E =E_{\text{MI-FNO}}-E_{\text{CCSD}}$, in kcal/mol) of the MI(3)-FNO approaches (MI(3)-FNO-UCCSD and MI(3)-FNO-CCSD) with respect to reference energy is plotted as a function of monotonically increasing virtual space size. The energy obtained with CCSD using full MO space is used as a reference energy. Plots are obtained for BeH$_\text{2}$. The area shaded in grey indicates where the results are within chemical accuracy from the reference energy.}
	\label{fig:Energies_BeH2}
\end{figure*}

Figure~\ref{fig:Energies_BeH2} shows how the total energy of BeH$_\text{2}$, using MI(2)-FNO-UCCSD, behaves as a function of the number of virtual orbitals. The plot shows the difference between the MI(2)-FNO-UCCSD and parent CCSD energies. The area filled in grey shows the region where deviations are within chemical accuracy. We see that the MI(2)-FNO-UCCSD values approach the reference CCSD energy as the number of virtual orbitals increases. When the number of virtual orbitals is seven, the difference from the reference energy becomes as small as 0.000860 hartrees, or 0.54 kcal/mol, showing that  chemical accuracy has been reached. For this calculation, there are three 16-qubit, one-body increments (one occupied and seven virtual orbitals) and three 18-qubit, two-body increments (two occupied and seven virtual orbitals).  When the number of  virtual orbitals is seven, we are able to discard 14 virtual orbitals. As the original problem requires 48 qubits without PD, this is a large reduction in quantum resources. We believe that the present MI-FNO framework can help accelerate the practical application of NISQ hardware in quantum chemistry simulations.

We do not run VQE calculations beyond 16 and 18 qubits because storing the exact state vector of the system in a classical device becomes challenging.To estimate the convergence behaviour of the energy as a function of the virtual space beyond 18 qubits, we first explore whether the MI(\(n\))-FNO-CCSD approach can be used to extrapolate the \mbox{MI(\(n\))-FNO-UCCSD} energies for the BeH$_\text{2}$ molecule. Varying the number of virtual orbitals from one to seven,  we confirm that the convergence behaviour of the MI(\(n\))-FNO-UCCSD and MI(\(n\))-FNO-CCSD approaches closely resemble each other. We then extend the MI(\(n\))-FNO-CCSD calculations to the maximum number of virtual orbitals (21 in the present setup) to gain an understanding of the convergence of the MI(2)-FNO-UCCSD energies. Based on this extrapolation, we find that MI(2)-FNO-UCCSD can provide chemically accurate results when the number of virtual orbitals is larger than seven.\\

\begin{figure*}[b]
	{\includegraphics[width=.48\linewidth]{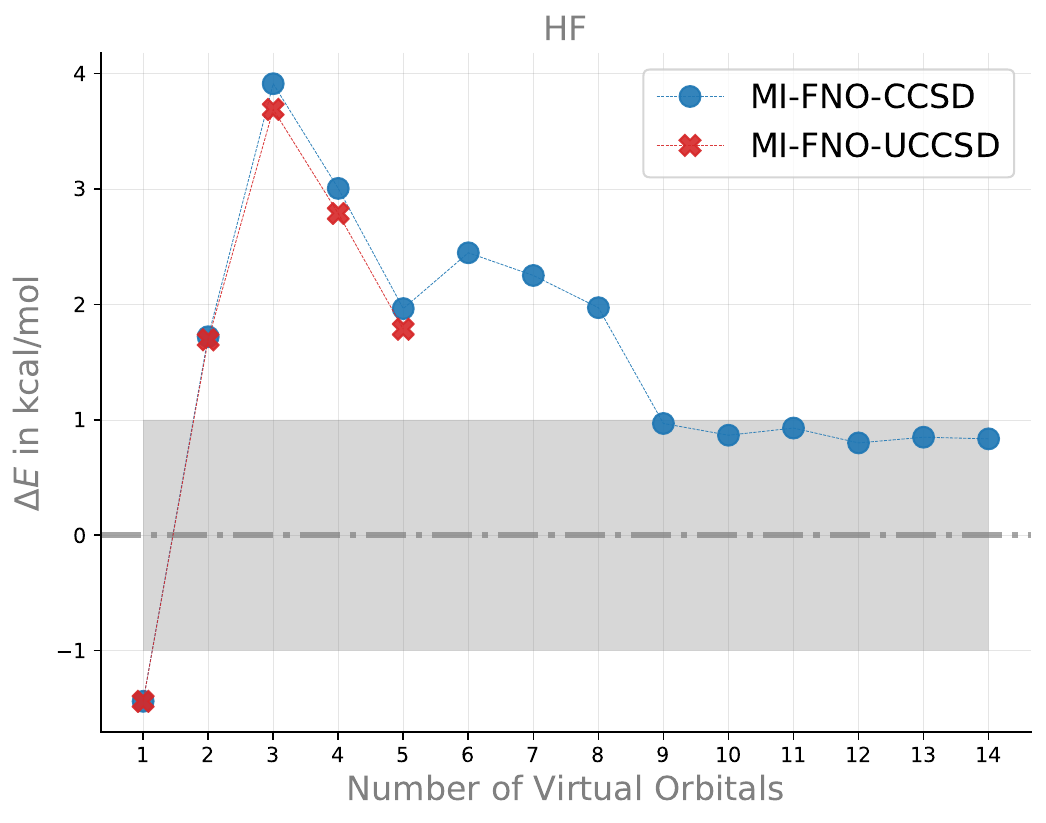} }\hfill
	\includegraphics[width=.48\linewidth]{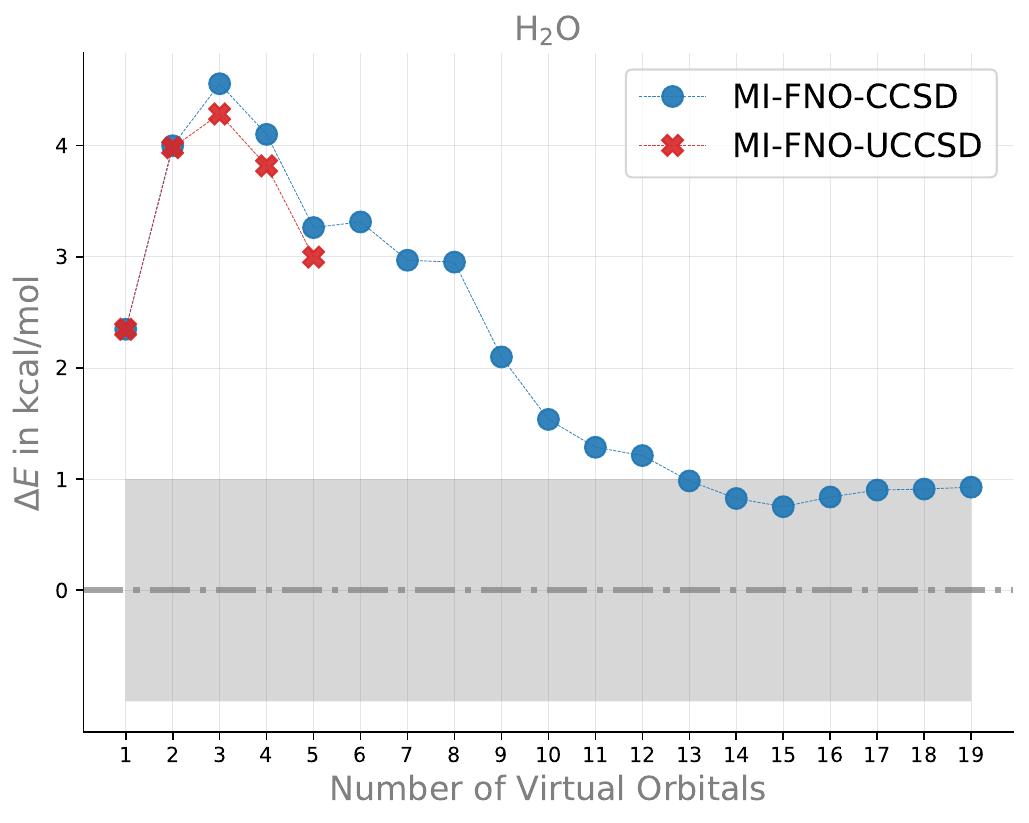} \\
	\vspace{5mm}
	
	\includegraphics[width=.48\linewidth]{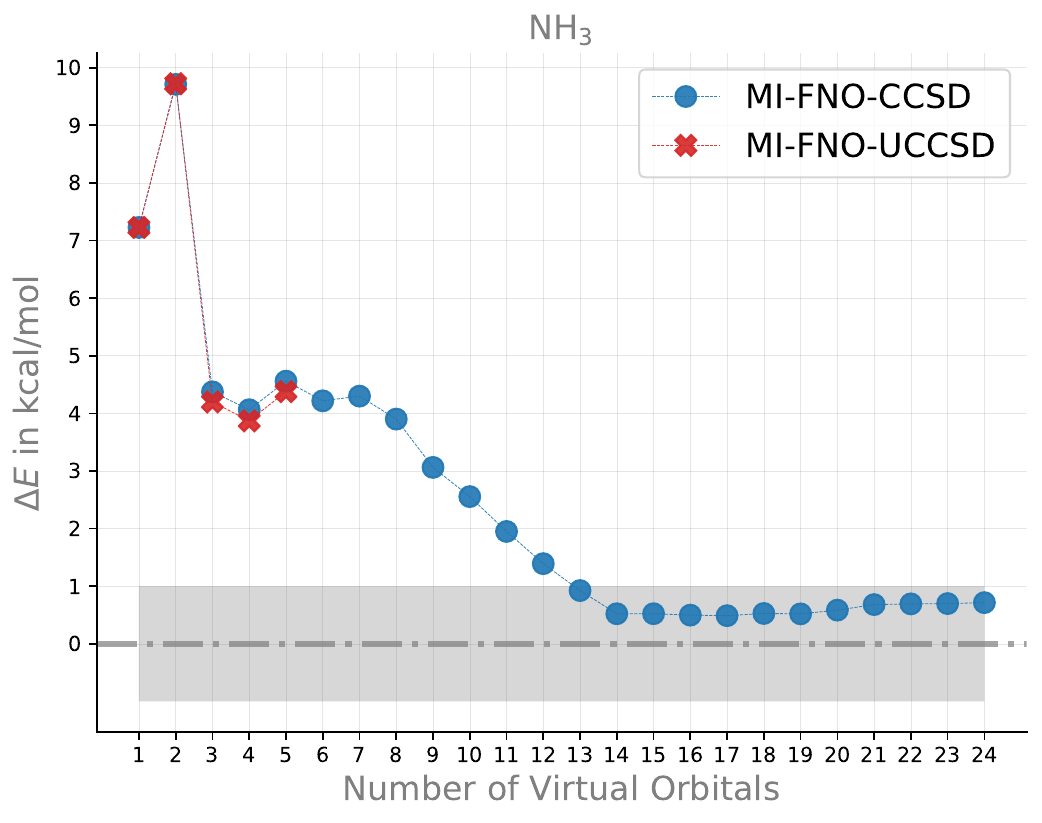}\hfill
	\includegraphics[width=.48\linewidth]{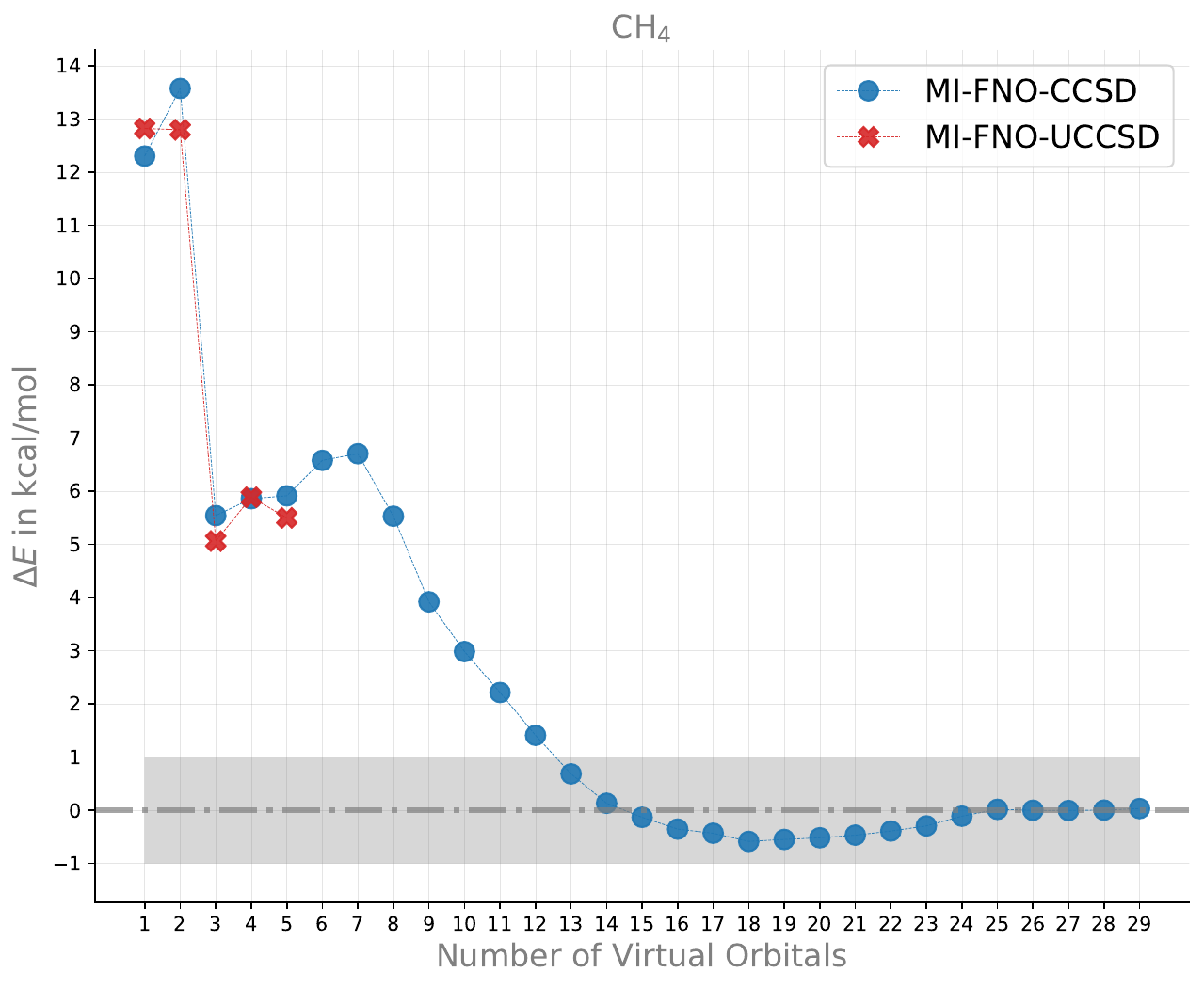}
	\caption{Energy deviation ($\Delta E =E_{\text{MI-FNO}}-E_{\text{CCSD}}$, in kcal/mol) of the MI(3)-FNO approaches (MI(3)-FNO-UCCSD and MI(3)-FNO-CCSD) with respect to reference energy is plotted as a function of monotonically increasing virtual space size. The energy obtained with CCSD using full MO space is used as a reference energy. Plots are obtained for HF, H$_\text{2}$O, NH$_\text{3}$, and CH$_\text{4}$. The area shaded in grey indicates where the results are within chemical accuracy from the reference energy.}
	\label{fig:Energies_}
\end{figure*}

Figure~\ref{fig:Energies_} contains similar information for  HF, H$_\text{2}$O,  NH$_\text{3}$, and CH$_\text{4}$. The \mbox{MI-FNO-UCCSD} energy profile for our test molecule closely follows that of MI-FNO-CCSD. As UCCSD and CCSD are fundamentally different theories, an exact equivalence between the results obtained from applying UCCSD and CCSD should not be expected. This is because the antisymmetric cluster operator in UCCSD also includes de-excitation operators and is solved by minimizing an energy functional variationally, while the CCSD ansatz includes only excitation operators and involves the solution of a set of non-linear, projected residual equations. However, for the weakly correlated hydrides considered in this work, we would expect UCCSD to give very similar results to CCSD near the equilibrium geometry.

The difference between the MI-FNO-UCCSD and MI-FNO-CCSD energies is just a fraction of kcal/mol. When the virtual space is smaller than the occupied space (i.e., $\le 3)$, the behaviour of the energy convergence is not very smooth and not monotonically decreasing.
Large errors (reaching a maximum of 12.8 kcal/mol for CH$_\text{4}$, 9.7 kcal/mol for NH$_\text{3}$, 9.2 kcal/mol for BeH$_\text{2}$, 4.3 kcal/mol for H$_\text{2}$O, and 4 kcal/mol for HF) are also associated with the smaller virtual space. When the virtual space become larger than the occupied space, the energy profile of the MI-FNO approach systematically converges to within chemical accuracy to the CCSD energy of the full molecular space. When all of the virtual space is taken into consideration, the MI-FNO-CCSD results show excellent convergence with respect to CCSD, while for NH$_\text{3}$, H$_\text{2}$O, and HF errors of approximately 1 kcal/mol are observed. 

The degree of the error may be mitigated by including higher-order body increments, so we examine the inclusion of four-body terms in the calculation (i.e., MI(4)-FNO-CCSD) as shown in Appendix~\ref{appendix:appA} in Figure~\ref{fig:Energies} (A), (B), (C), and (D). In all cases considered, the inclusion of the four-body terms in the MI(4)-FNO-CCSD calculation improves the energy convergence towards the reference energy, and the energy at the point where the number of virtual orbitals is equal to the full virtual space differs from the reference energy by only around $5\times 10^{-5}$ hartrees or less. The reason why we observe highly accurate results with the MI(4)-FNO-CCSD approach for these molecules may be because they are all 10-electron systems and the contribution from the core electrons is usually very small when the basis sets have no core-polarization functions, such as in the cc-pVDZ basis used here. We plan to investigate the impact of  higher-order body increments  (e.g., three-body, four-body, five-body) on the convergence behaviour of the MI-FNO approach by targeting larger molecular systems.

We also speculate that the error in the MI-FNO approach is due to the fact that the occupied orbitals are not spatially localized, and therefore the decomposition of the occupied space into a smaller space, based on the method of increments, causes the residual error. We observe that using Foster--Boys localization~\cite{Foster:1960aa} improves the energy convergence of \mbox{MI(3)-FNO-CCSD} towards the reference energy. We plan to discuss this improvement in more detail in a future publication.

The accuracy of approximating full molecular space correlation energies using MI and conventional quantum chemistry approaches such as FCI and CC has been explored by various research groups.\cite{Stoll:1992aa, Modll:1997aa, Bezugly:2004aa, Stoll:2005aa, Friedrich:2007aa, Dahlke2007aa, Bytautus:2010aa, Gordon:2012ab, Mueller:2012aa, Richard:2012, Friedrich:2013a, Friedrich:2013b, Zhang:2013ab, Voloshina:2014aa, Anacker:2016, Lao:2016aa, Fiedler:2017, Eriksen:2017aa, Boschen:2017aa, Eriksen:2018, Fertitta:2018aa, Zimmerman:2019aa, Eriksen:2019a, Eriksen:2019b} 
To validate the implementation of the MI($n$) approach for our test molecules,
the total energies of BeH$_\text{2}$, CH$_\text{4}$, NH$_\text{3}$, H$_\text{2}$O, and HF are calculated using the MI(2)-CCSD, MI(3)-CCSD, and MI(4)-CCSD methods  with no virtual space truncation. A comparison of the total energies using MI and the full virtual space  with conventional CCSD energies is presented in Appendix~\ref{appendix:appA} in Table~\ref{tab:MI_total_ene}. As discussed in the same section, an accuracy of 1 kcal/mol is achieved for these molecules using the MI(3) expansion. To validate the MI approach, in conjunction with a virtual space truncation based on the FNO approach (MI-FNO-CCSD), the conventional  criterion of 99{\%} occupancy for the virtual space truncation is used. The total energies calculated using MI-FNO-CCSD, and their difference from the reference CCSD total energies, are listed in the second section of Appendix~\ref{appendix:appA} in Table~\ref{tab:MI_FNO_total_ene}.  As discussed in that section, chemical accuracy is achieved for the total energies of the test molecules using MI-FNO-CCSD and a 99{\%} FNO population percentage criterion. By plotting the cumulative FNO occupancy percentage as a function of the number of virtual orbitals (see Figure~\ref{fig:FNO}), the FNO procedure is quite effective at exploiting the sparsity in the virtual space. The FNO procedure produces a virtual space that is compact and becomes even more effective at compressing the size of this space as the size of the basis is increased.

\subsection{Future Outlook of the MI-FNO Framework}

In order to use VQE to test continuously evolving NISQ devices in chemistry applications, one should employ a framework such as MI-FNO which provides systematically improvable results with increasing quantum computational resources. To demonstrate that the \mbox{MI-FNO} approach is an effective framework for systematically reducing quantum resources for applications of VQE, we provide estimates for the qubit count for an industrially relevant catalyst molecule.

\begin{figure*}[h]
	\includegraphics[scale=0.2]{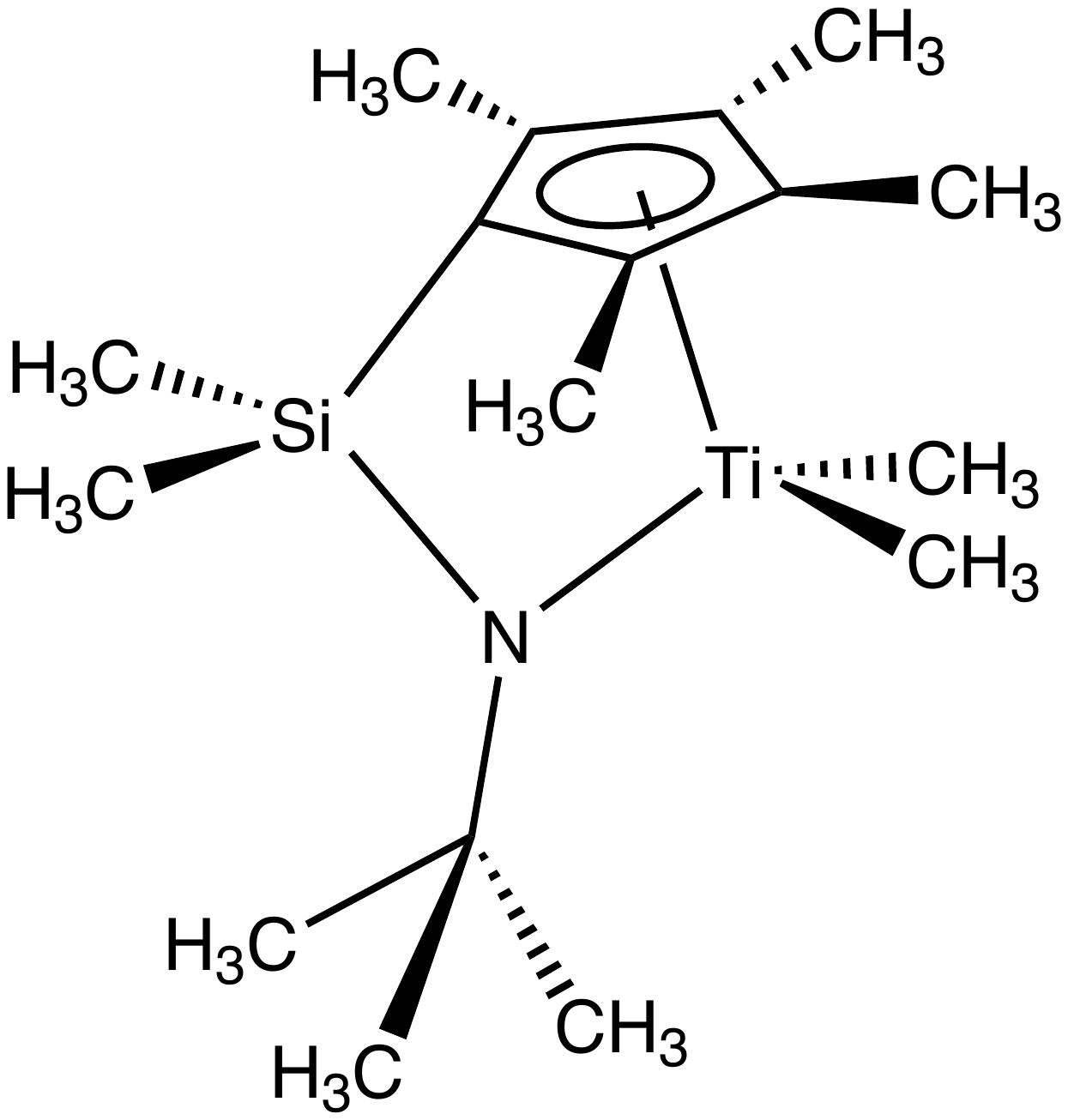}
	\caption{Molecular structure of a CGC catalyst. The image has been adapted from Figure 1 of Ref. 87.}
	\label{fig:CGC1}
\end{figure*}

\begin{figure*}[b]
	\includegraphics[scale=0.4]{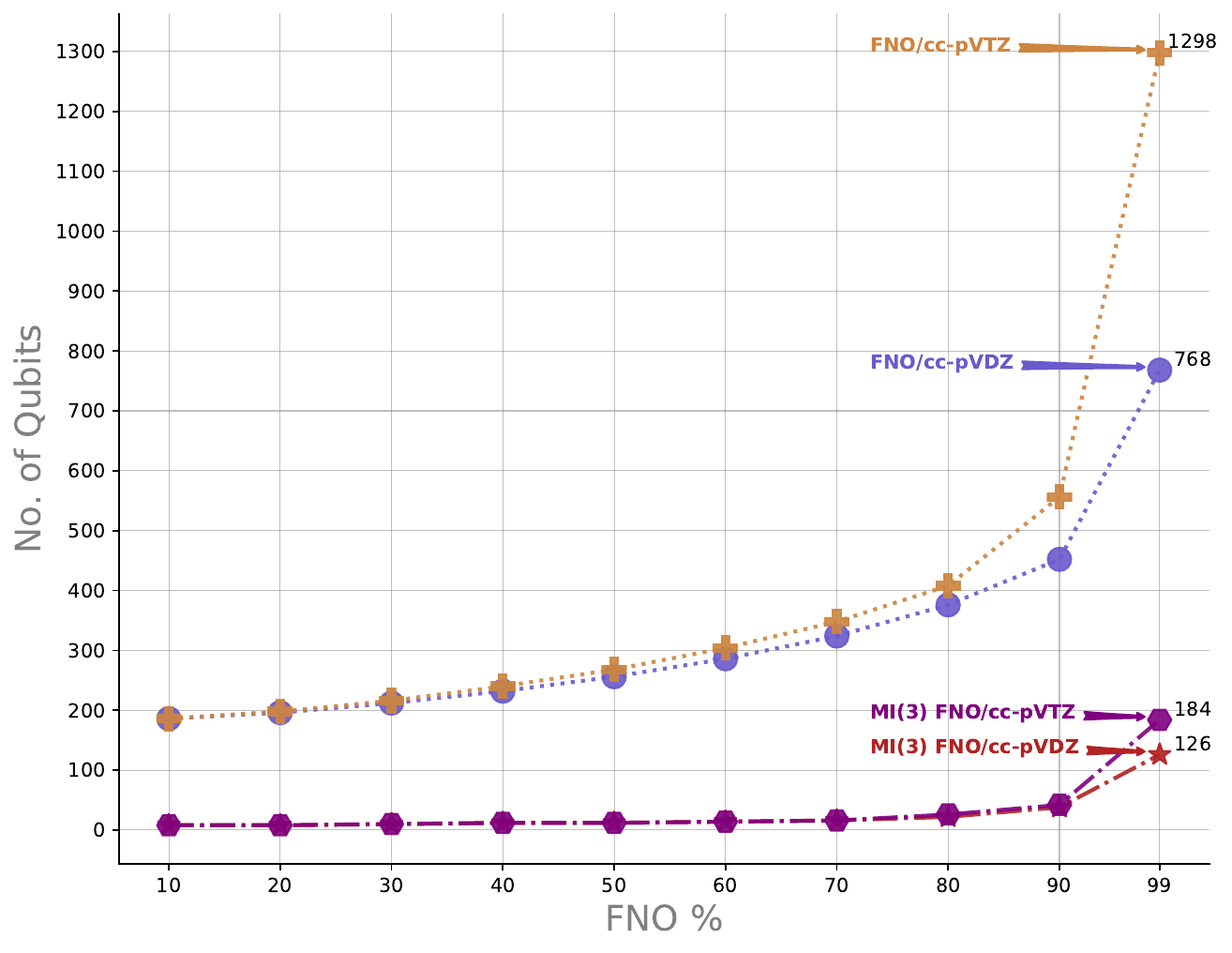}
	\caption{Qubit count estimates for a CGC catalyst. The molecule has 89 occupied orbitals. It has, respectively, 389 and 1015 virtual orbitals for the cc-pVDZ and cc-pVTZ basis sets. The y-axis represents the number of qubits needed to perform the methods, while the x-axis represents the FNO\% that is used to truncate the virtual space. The methods considered in the plots are FNO and MI(3)-FNO. In FNO, only the virtual space is truncated while in MI(3)-FNO, both the occupied and virtual spaces are truncated. Numbers on the right (1298, 768, 184, and 126) represent the number of qubits needed when the virtual space is truncated using a 99\% FNO population percentage.}
	\label{fig:Quantum_Resources}
\end{figure*}

We provide these qubit counts for the CGC(1) catalyst from Figure 1 in Arriola et al.~\cite{Arriola:2007aa} (see Figure~\ref{fig:CGC1} above) as an early indication of the efficacy of the MI($n$)-FNO approach.
The crystal structure of CGC(1) is obtained from Ref.~\citenum{Arriola:2007aa}, and the cc-pVDZ and the cc-pVTZ basis sets are used for the quantum resource estimations.
In Figure \ref{fig:Quantum_Resources}, we summarize the qubit count estimations when the FNO virtual space truncation is applied on its own and when the MI(3)-FNO approach is applied.

To estimate the number of qubits for the FNO virtual space truncation, we add the number of FNO virtual orbitals $N_\text{v}$ after truncation to the number of occupied orbitals $N_\text{occ}$ in the system. 
The number of virtual orbitals $N_\text{v}$ in the FNO approach is determined based on a given percentage of FNO occupancy for the full molecular system. So, the total qubit count is obtained using the expression $2(N_\text{v} + N_\text{occ})$. 
In the qubit count estimation for the \mbox{MI(3)-FNO} approach, the maximum number of occupied orbitals is three, corresponding to the three-body increments in the MI(3) expansion. Hence, the total qubit count in Figure~\ref{fig:Quantum_Resources} is obtained using the expression \(2(N_\text{rv} + 3)\), where $N_\text{rv}$ is obtained using our FNO procedure with the MBPT2 density that has only 3 occupied orbitals as active orbitals. 
This is a rough estimate of the number of qubits needed for the MI(3)-FNO approach as we do not scan all the increments. The number reported in Figure \ref{fig:Quantum_Resources} for the MI-FNO approach corresponds to the increment that has three occupied orbitals: HOMO, HOMO-1, and HOMO-2.

The MI(3) approach, without truncation of the virtual space, reduces the qubit requirements by 172 for a CGC catalyst.
Further reduction in qubit counts is achieved by employing an FNO population percentage (i.e., 99\%) to truncate the virtual space. The MI(3)-FNO approach significantly reduces the qubit requirements.
The UCCSD with the full molecular space calculation needs 956 and 2208 qubits, respectively, for the cc-pVDZ and cc-pVTZ basis sets, whereas FNO-UCCSD with a 99\% FNO population percentage needs 768 and 1298 qubits, respectively, for the basis sets. 
This approach leads to a virtual space reduction of 25\% and 45\%, respectively, for the cc-pVDZ and cc-pVTZ basis sets (see Figures \ref{fig:cgc_FNO_DZ} and \ref{fig:cgc_FNO_TZ} and the rest of Appendix~\ref{appendix:appC} for more details). 
Given the finding in the second section of Appendix~\ref{appendix:appA} (see Table~\ref{tab:MI_FNO_total_ene}) that the MI-FNO approach with a 99\% FNO occupancy can produce very accurate results for our test molecules, we speculate that it is possible to obtain results of similar quality for larger molecules. The MI(3)-FNO-UCCSD approach with a 99\% FNO population percentage needs 126 and 184 qubits, respectively, for the cc-pVDZ and cc-pVTZ basis sets.
This approach leads to a virtual space reduction of 85\% and 90\%, respectively, for the basis sets (see Figure \ref{fig:cgc_MI_DZ} and \ref{fig:cgc_MI_TZ} and the rest of Appendix~\ref{appendix:appC} for more details). 
The number of qubits is drastically decreased by truncating the virtual space with a smaller FNO population percentage. Also, for a smaller FNO percentage, the number of qubits required does cause an appreciable change for both basis sets. Comparison of the qubit count between the FNO and MI(3)-FNO approaches demonstrates that the MI(3)-FNO approach has a much smaller qubit requirement than the FNO approach. 
We believe that the MI-FNO approach is quite beneficial for calculating electron correlation energies for larger molecular systems, where one must deal with many occupied orbitals. It is needless to mention that the number of increments in the MI-FNO approach will become larger for a system with a large occupied space.
For example, a CGC catalyst has about 117,569 increments in the three-body expansion. Furthermore, an effective screening procedure such as the distance-, energy-, and domain-based approaches~\cite{Friedrich:2007aa, Anacker2016} can be implemented to reduce the number of increments while maintaining the chemical accuracy of the calculation. 
It would be interesting to explore what the minimum quantum resource requirements would be for obtaining chemically accurate total energies for these large molecular systems after implementing a highly parallel framework.

\section{Conclusion} \label{sec:conclusion}
Quantum computing is an alternative computational paradigm with the potential to accelerate the materials innovation process, thereby reducing the time to new discoveries.
In the era of NISQ devices, VQE  has emerged as a promising algorithm for characterizing the usefulness of NISQ devices for quantum chemistry applications. However, NISQ devices have to overcome many challenges before they become useful for chemical applications.
For example, some of the main bottlenecks are the design of scalable physical quantum states with long coherence times and fast gate operations with low error rates. Efforts have been made to mitigate issues arising from large quantum resource requirements, coherence and run times of quantum circuits, noisy gate operations, measurements of energy, and the classical optimization of ansatz parameters.

In the present study, we have focused on reducing the problem size as a strategy for utilizing current and near-term NISQ devices for the simulation of molecular systems. We believe a reduction in qubit count to be essential in helping to advance the timeline for applications of quantum computing in materials science.
At the same time, it could provide opportunities for further characterization of the usefulness of NISQ devices for quantum chemistry simulations by allowing hardware experiments to be conducted on smaller, yet more realistic, chemistry problems. We have described a novel framework for the systematic reduction of both the occupied and virtual spaces of molecular systems. Our MI-FNO approach distributes the occupied orbitals among $n$-body increments, based on the many-body expansion of the correlation energy in terms of the occupied orbital space, while a scalable framework for the virtual orbital space is created by using the FNO approach.

As a demonstration of the applicability of the MI-FNO approach, we used VQE in combination with the UCCSD ansatz.
We examined its accuracy and feasibility by studying small molecules, namely, BeH$_\text{2}$, CH$_\text{4}$, NH$_\text{3}$, H$_\text{2}$O, and HF, in a cc-pVDZ basis set.
We observed that the MI-FNO approach can achieve chemical accuracy by significantly reducing both the number of qubits and the number of gate operations, which suggests that it can be used to build a scalable quantum chemistry simulation platform that effectively utilizes quantum hardware.
Furthermore, as an early demonstration of the efficacy of this approach for larger molecules, we have presented qubit count estimations for a titanium-metal-based CGC catalyst that has relevance in the large-scale polymerization of $\alpha$-olefin.
We found that by employing a modest truncation of the virtual space using a 99\% FNO occupancy, a significant reduction in the qubit requirements can be achieved.

\section*{Acknowledgements}
This work was supported as part of a joint development agreement between Dow and 1QBit. We are grateful to Alejandro Garza and Peter Margl from Dow for technical discussions and guidance regarding industrial chemistry use cases and applications, and to Paul M. Zimmerman at the University of Michigan for technical discussions. The authors thank Marko Bucyk at 1QBit for reviewing and editing the manuscript.

\section*{Data Availability}
The data that supports the findings of this study are available within the article.

\newpage
%\section*{References}
\bibliographystyle{aipnum4-1}
\bibliography{my_bib_fix}

\newpage
\appendix
\counterwithin{figure}{section}
\counterwithin{table}{section}
\section*{Appendices}
\section{Numerical Validation of the MI(\(n\))-FNO Implementation \label{appendix:appA}}
\subsection{Accuracy of the Method of Increments without Truncation of the Virtual Space} \label{appendix:MI_accuracy}
In the MI-FNO approach, two sources of approximation are used. The correlation energy is approximated by using the MBE and the virtual space is truncated using the FNO procedure. 
To validate the MI($n$)-FNO approach, we first examine the accuracy of the energy calculations using the method of increments, with no virtual space truncation. 
The total energies of BeH$_\text{2}$, CH$_\text{4}$, NH$_\text{3}$, H$_\text{2}$O, and HF are calculated using the MI(2)-CCSD, MI(3)-CCSD, and MI(4)-CCSD methods. The total energies computed using MI and the full virtual space are then compared against conventional full molecular space CCSD energies.

\begin{table} [h]
	\caption {Total energy values (hartrees) and the difference from the conventional CCSD values using the MI(\(n\))-CCSD approach. The differences are shown in parentheses. The calculated results for the many-body expansion truncated up to \(n\) = 2-, 3-, and 4-body increments are listed.} \label{tab:MI_total_ene}
	\begin{tabular}{ c | c | c c | c c | c c}
		\multicolumn{1}{c|}{} & \multicolumn{1}{c|}{CCSD} & \multicolumn{2}{c|}{MI(2)-CCSD} & \multicolumn{2}{c|}{MI(3)-CCSD} & \multicolumn{2}{c}{MI(4)-CCSD}\\
		\hline
		BeH$_2$ & -15.835746 & -15.835806 & (-0.000060) & -15.835746 & (0.000000) & – & –\\
		CH$_4$ & -40.385951 & -40.392778 & (-0.006827) & -40.385350 & (0.000601) & -40.385952 & (-0.000001) \\
		NH$_3$ & -56.400579 & -56.412272 & (-0.011693) & -56.399440 & (0.001140) &  -56.400581 & (-0.000002) \\
		H$_2$O & -76.240099 & -76.254995 & (-0.014896) & -76.238622 & (0.001478) &  -76.240102 & (-0.000003) \\
		HF & -100.228154 & -100.242731 & (-0.014577) & -100.226824 & (0.001331) & -100.228157 & (-0.000003) \\
		\hline
	\end{tabular}
\end{table}
The total energies using the MI methods and the comparison of the energies with the CCSD energies are listed in Table \ref{tab:MI_total_ene}. We achieve chemical accuracy (0.0015 hartrees or 1.0 kcal/mol) with respect to the conventional CCSD value for the six-electron system BeH$_\text{2}$ using the MI(2) approach. The MI(3) value agrees with the parent CCSD value, because they are equivalent for this six-electron system. Hence, we find that MI(2) is sufficiently accurate for performing calculations on BeH$_\text{2}$. For the other four systems, each of which contains 10 electrons, we observe a  relatively large error using the MI(2) expansion. The CH$_\text{4}$ molecule exhibits the smallest error, $-0.006827$ hartrees ($4.28$ kcal/mol), which is over four times larger than the target value of $1.0$ kcal/mol needed for chemical accuracy. In contrast, with the MI(3) approach, we achieve chemical accuracy in the total energies for all molecules considered. The largest error was observed for the total energy of H$_\text{2}$O, which is $0.001478$ hartrees ($0.93$ kcal/mol) larger than the CCSD value. The error becomes less than \mbox{$5.0 \times 10^{-6}$ hartrees} when using MI(4) for these 10-electron systems. The accuracy of the MI(2) expansion is not sufficient for achieving chemical accuracy for the 10-electron systems considered in this work. 

The number of occupied orbitals in CCSD calculations is reduced by decomposing the original problem into subproblems (increments) using the MI expansion. For BeH$_\text{2}$, the MI(2) calculation includes only two occupied orbitals, while three occupied orbitals are used in the CCSD calculation. For the other 10-electron systems, the MI(3) and MI(4) calculations include three and four occupied orbitals, respectively, while full CCSD calculation includes five occupied orbitals. In this study, we use relatively small-sized molecules; thus, the reduction of occupied orbitals is small. However, as we show later in this section, if we apply MI methods to larger-sized systems, we achieve a large reduction in the number of occupied orbitals. The MI method has the potential to recover accurate total energies while reducing computational costs.   

\subsection{Accuracy of the Method of Increments with Truncation of the Virtual Space}\label{appendix:MI_FNO_accuracy}

To investigate the accuracy of molecular energy calculations using the  MI approach, in conjunction with virtual space truncation based on the FNO approach (MI-FNO-CCSD), we choose the criterion of virtual orbital selection using a population percentage of 99{\%}. 

\begin{table} [tbh]
	\caption {Total energy values (hartrees) and the difference from the conventional CCSD values using the  MI(\(n\))-FNO-CCSD approach. The differences are shown in parentheses. We employ an FNO population of 99{\%} to determine the size of the virtual space.} \label{tab:MI_FNO_total_ene}
	\renewcommand{\arraystretch}{1.2}
	\begin{tabular}{ c | c | c c | c c}
		\multicolumn{1}{c|}{} & \multicolumn{1}{c|}{CCSD} & \multicolumn{2}{c|}{MI-CCSD} & \multicolumn{2}{c}{MI-FNO-CCSD} \\
		\hline
		BeH$_2$ MI(2) & -15.835746 & -15.835806 & (-0.000060) & -15.836066 & (-0.000320)  \\
		CH$_4$ MI(3) & -40.385951 & -40.385350 & (0.000601) & -40.385716 & (0.000235)  \\
		CH$_4$ MI(4) & -40.385951 & -40.385952 & (-0.000001) & -40.386196 & (-0.000245)  \\
		NH$_3$ MI(3) & -56.400579 & -56.399440 & (0.001140) & -56.399468 & (0.001111)  \\
		NH$_3$ MI(4) & -56.400579 & -56.400581 & (-0.000002) & -56.400693 & (-0.000114) \\
		H$_2$O MI(3) & -76.240099 & -76.238622 & (0.001478) & -76.238514 & (0.001585)\\
		H$_2$O MI(4) & -76.240099 & -76.240102 & (-0.000003) & -76.239986 & (0.000113) \\
		HF MI(3) & -100.228154 & -100.226824 & (0.001331) & -100.226893 & (0.001261) \\
		HF MI(4) & -100.228154 & -100.228157 & (-0.000003) & -100.227998 & (0.000156)  \\
		\hline
	\end{tabular}
\end{table}

The total energies calculated using MI-FNO-CCSD, and their difference from the reference CCSD total energies, are listed in Table~\ref{tab:MI_FNO_total_ene}. Chemical accuracy is achieved for the total energies calculated with all the MI-FNO-CCSD approaches. The MI(3)-FNO-CCSD calculation performed on the H$_2$O molecule exhibits the largest error of 0.001478 hartrees (0.99 kcal/mol). The FNO approach reduces the number of virtual orbitals for each subproblem in the MI expansion. For BeH$_2$, using the MI(2) expansion, the FNO method with the threshold of 99\% occupancy discards 17, five, and seven virtual orbitals from the three one-body increments, and five, seven, and six virtual orbitals from the three two-body increments, while the full problem of BeH$_2$ has 21 virtual orbitals. Therefore, the FNO method with a 99\% threshold is able to discard at least five virtual orbitals. For the other 10-electron systems with the MI(3) expansion, the FNO approach discards at least seven, five, three, and two virtual orbitals for the CH$_4$, NH$_3$, H$_2$O, and HF molecules, respectively. Again, we consider smaller systems in this work, so the reduction may not appear significant. However, we observe that the FNO virtual space truncation becomes more efficient as the virtual space becomes larger.

\begin{figure*}[tbp]
	\includegraphics[scale=0.5]{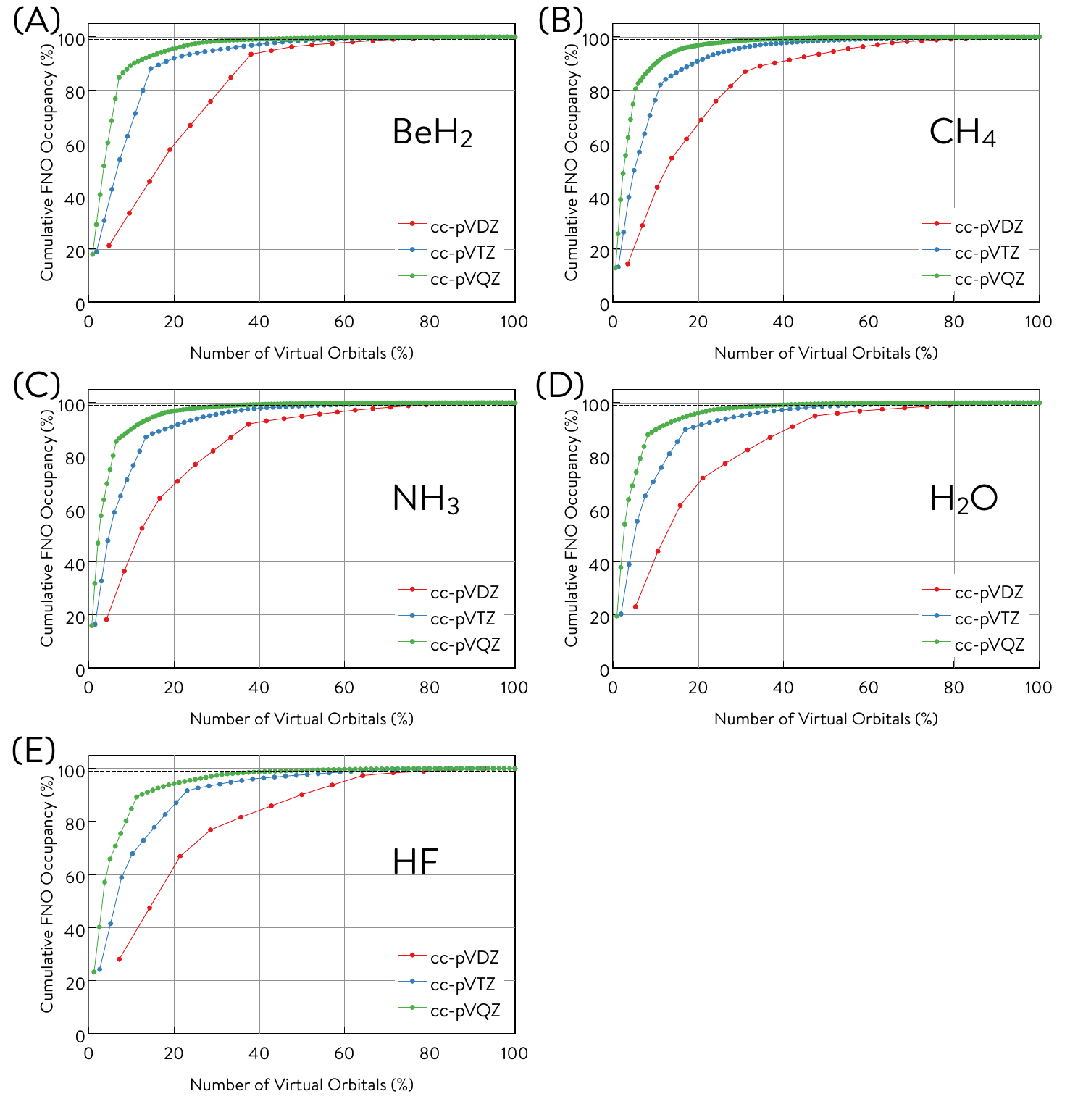}
	\caption{Cumulative FNO occupancy as a function of the number of virtual orbitals plotted for BeH$_2$, CH$_4$, NH$_3$,  H$_2$O, and HF.}
	\label{fig:FNO}
\end{figure*}

\begin{figure*}[tbp]
	\includegraphics[scale=0.45]{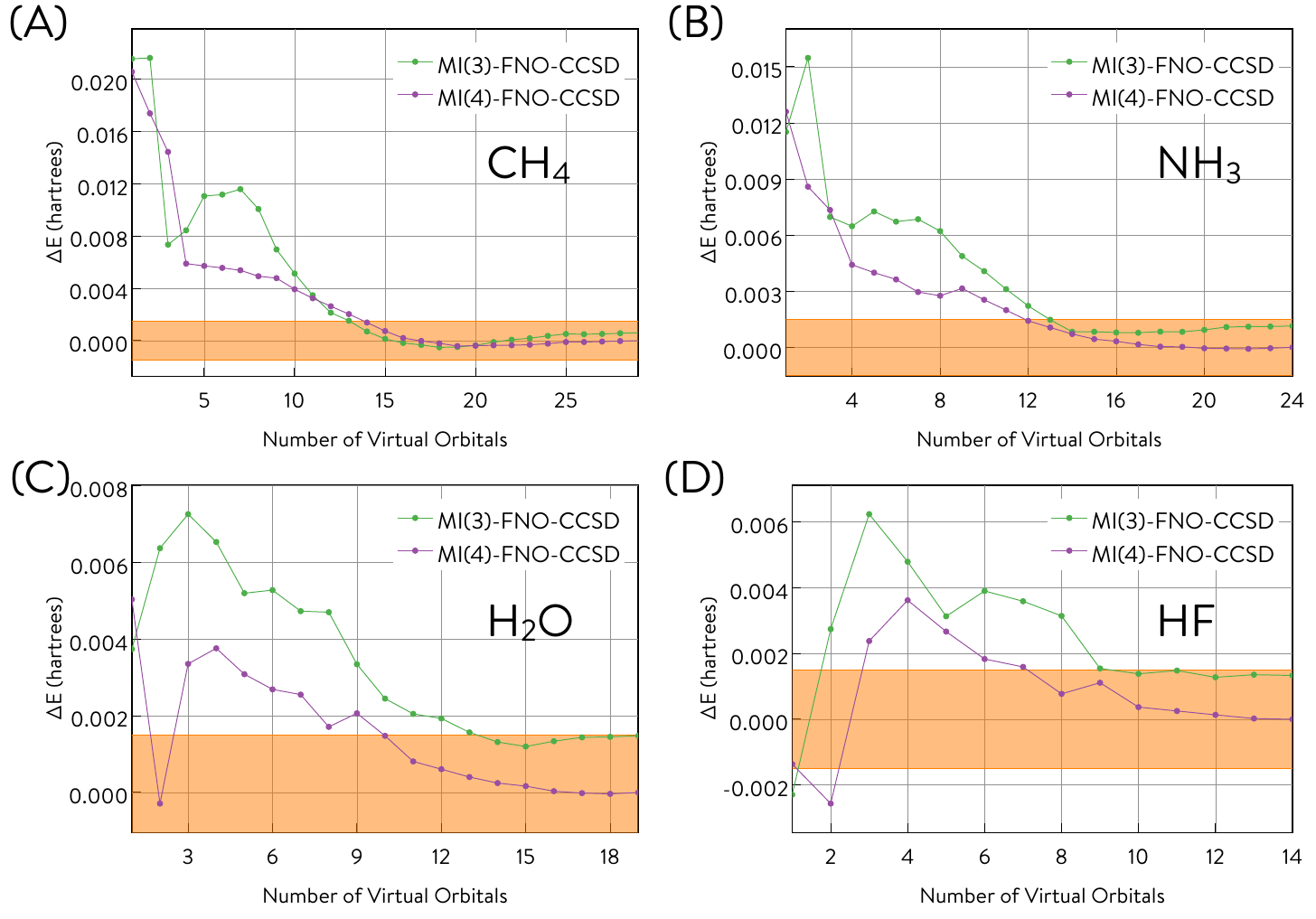}
	\caption{Energy deviation ($\Delta E =E_{\text{MI-FNO}}-E_{\text{CCSD}}$, in hartrees) of the MI($n$)-FNO approaches (MI(3)-FNO-CCSD and MI(4)-FNO-CCSD) with respect to the reference energy is plotted as a function of monotonically increasing virtual space size. The energy obtained with CCSD using the full MO space is used as the reference energy. Plots are obtained for CH$_\text{4}$, NH$_\text{3}$, H$_\text{2}$O, and HF. The area shaded in orange indicates where the results are within chemical accuracy to the reference energy.}
	\label{fig:Energies}
\end{figure*}

Figure~\ref{fig:FNO} shows the cumulative FNO occupancy percentage as a function of the number of virtual orbitals for the molecules we examine. The values on the horizontal axis represent the ratio of the number of virtual orbitals, calculated as ``the number of virtual orbitals that are used in the calculation'' divided by ``the total number of virtual orbitals of the system''. The dotted line shows the FNO occupancy of 99\%. The plots are obtained by running  FNO-CCSD calculations, not by using the MI-FNO-CCSD approach. As shown, the larger basis set reaches the 99\% FNO occupancy faster than the smaller basis sets. This means that the FNO truncation discards more virtual orbitals as the virtual space becomes larger. Therefore, if we were to apply MI-FNO-CCSD to larger-sized systems or employ larger basis sets, we would achieve not only a considerable reduction in the number of occupied orbitals but also a significant reduction in the number of virtual orbitals. We find that the MI-FNO-CCSD method accurately recovers the total molecular energies, while reducing the computational cost, in comparison with the CCSD approach.

The many-body expansion of the correlation energy can be truncated to three-body or four-body terms. To show the impact an $n$-body has on energy, a comparative study of energy convergence behaviour is shown in Figure~\ref{fig:Energies}. Deviation of the energy of MI(3)-FNO-CCSD and MI(4)-FNO-CCSD with respect to the energy of CCSD is plotted against the monotonically increasing virtual space. The four-body expansion recovers full CCSD energy well for the 10-electron systems.  

\clearpage

\section{Detailed Quantum Resource Estimation} \label{appendix:appB}
\begin{table}[h]
	\caption {Computational space for CCSD in terms of the number of electrons and molecular orbitals, and quantum computational space of UCCSD in terms of the number of qubits, one-qubit gates, and two-qubit gates.} \label{tab:quantum_computational_space}
	\renewcommand{\arraystretch}{1.1}
	\setlength{\tabcolsep}{0.5em}
	\begin{tabular}{ c | c  c | c  c  c }
		&		 \multicolumn{2}{c|}{CCSD}	&		 \multicolumn{3}{c}{UCCSD}			\\
		Molecules	&	Electrons	&	Mol. Orbitals	&	Qubits	&	 One-Qubit Gates	&	 Two-Qubit Gates	\\
		\hline
		BeH$_2$	&	6	&	24	&	48	&	7.32E+04	&	3.02E+05	\\
		HF	&	10	&	19	&	38	&	2.05E+05	&	6.58E+05	\\
		H$_2$O	&	10	&	24	&	48	&	2.41E+05	&	9.30E+05	\\
		NH$_3$	&	10	&	29	&	58	&	7.32E+05	&	3.37E+06	\\
		CH$_4$	&	10	&	34	&	68	&	1.73E+06	&	9.48E+06	\\
		\hline
	\end{tabular}

\end{table}
\begin{table} [h]
	\caption {Number of qubits required to obtain chemical accuracy using the MI($n$)-FNO-UCCSD approach. The number of qubits is estimated based on the energies obtained with the corresponding MI($n$)-FNO-CCSD approach for CH$_4$, NH$_3$, H$_2$O, and HF. The numbers in parentheses indicate the number of increments the MI($n$) approach generates.} \label{tab:number_of_qubits}
	\begin{tabular}{ c | c c c c}
		& UCCSD & MI(2)-FNO-UCCSD & MI(3)-FNO-UCCSD & MI(4)-FNO-UCCSD \\
		\hline
		BeH$_2$ & 48 & 18 (6) & – & –\\
		CH$_4$ & 68 & – & 32 (25) &  36 (30) \\
		NH$_3$ & 58 & – & 32 (25) &  32 (30)\\
		H$_2$O & 48 & – & 34 (25) &  28 (30)\\
		HF & 38 & – & 26 (25) & 24 (30) \\
		\hline
	\end{tabular}
\end{table}

\begin{figure*}[tbp]
	\includegraphics[scale=0.38]{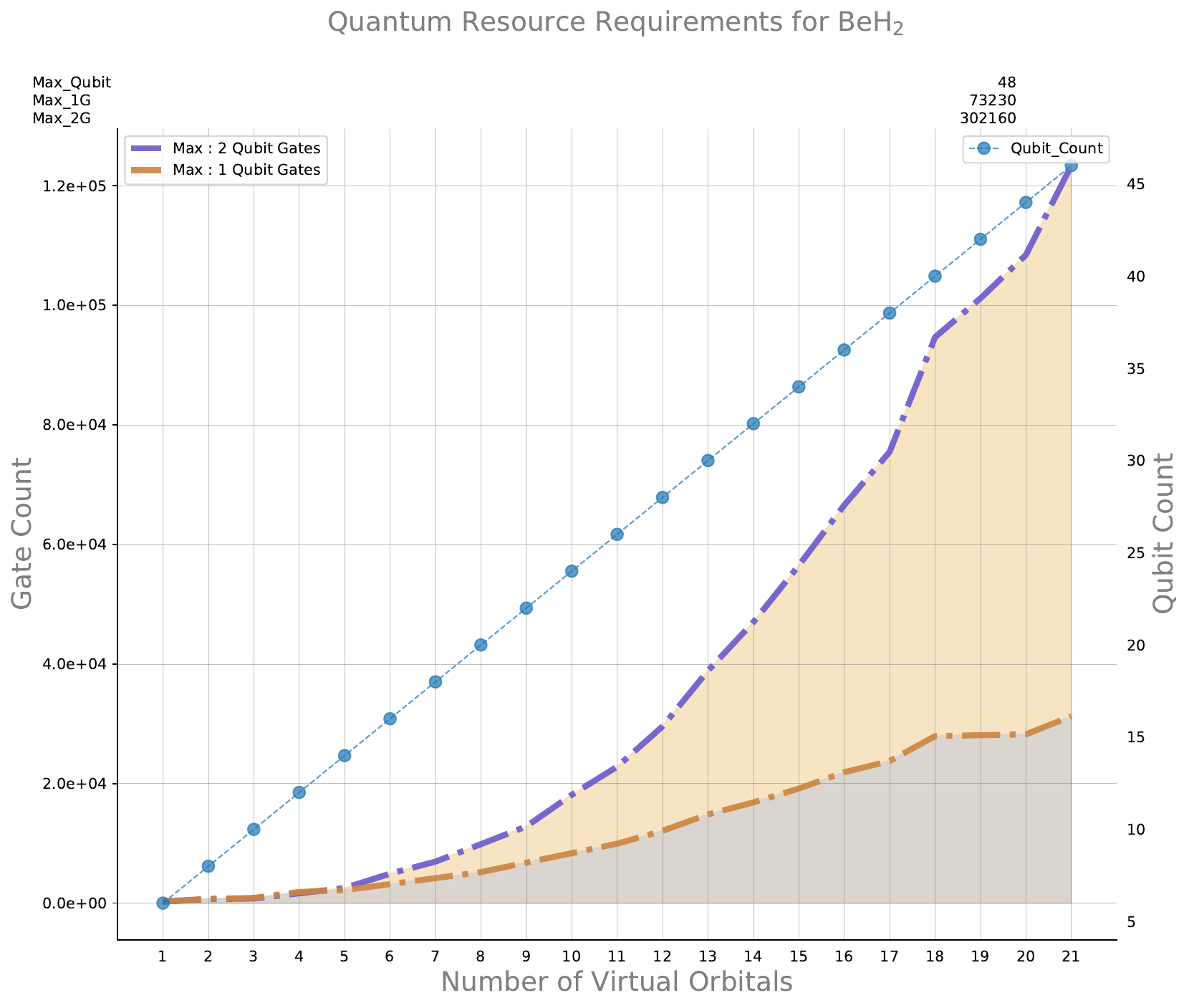}
	\caption{Quantum resources required to prepare the quantum state of the subsystem of BeH$_2$ as a function of the number of virtual orbitals}
	\label{fig:Quantum_Resources_BeH2}
\end{figure*}

\begin{figure*}[tbp]
	\includegraphics[scale=0.38]{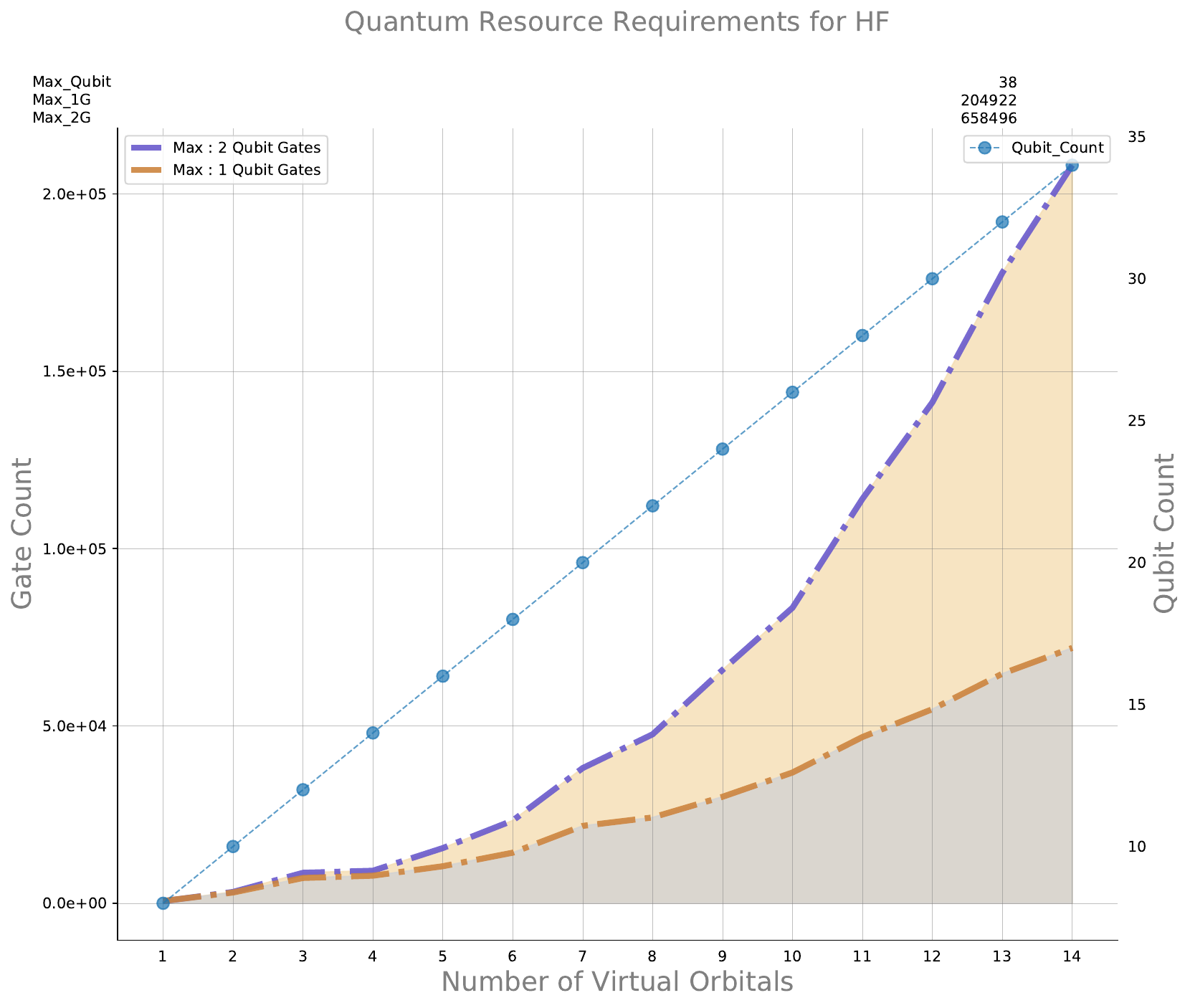}
	\caption{Quantum resources required to prepare the quantum state of the subsystem of HF as a function of the number of virtual orbitals}
	\label{fig:Quantum_Resources_HF}
\end{figure*}

\begin{figure*}[tbp]
	\includegraphics[scale=0.38]{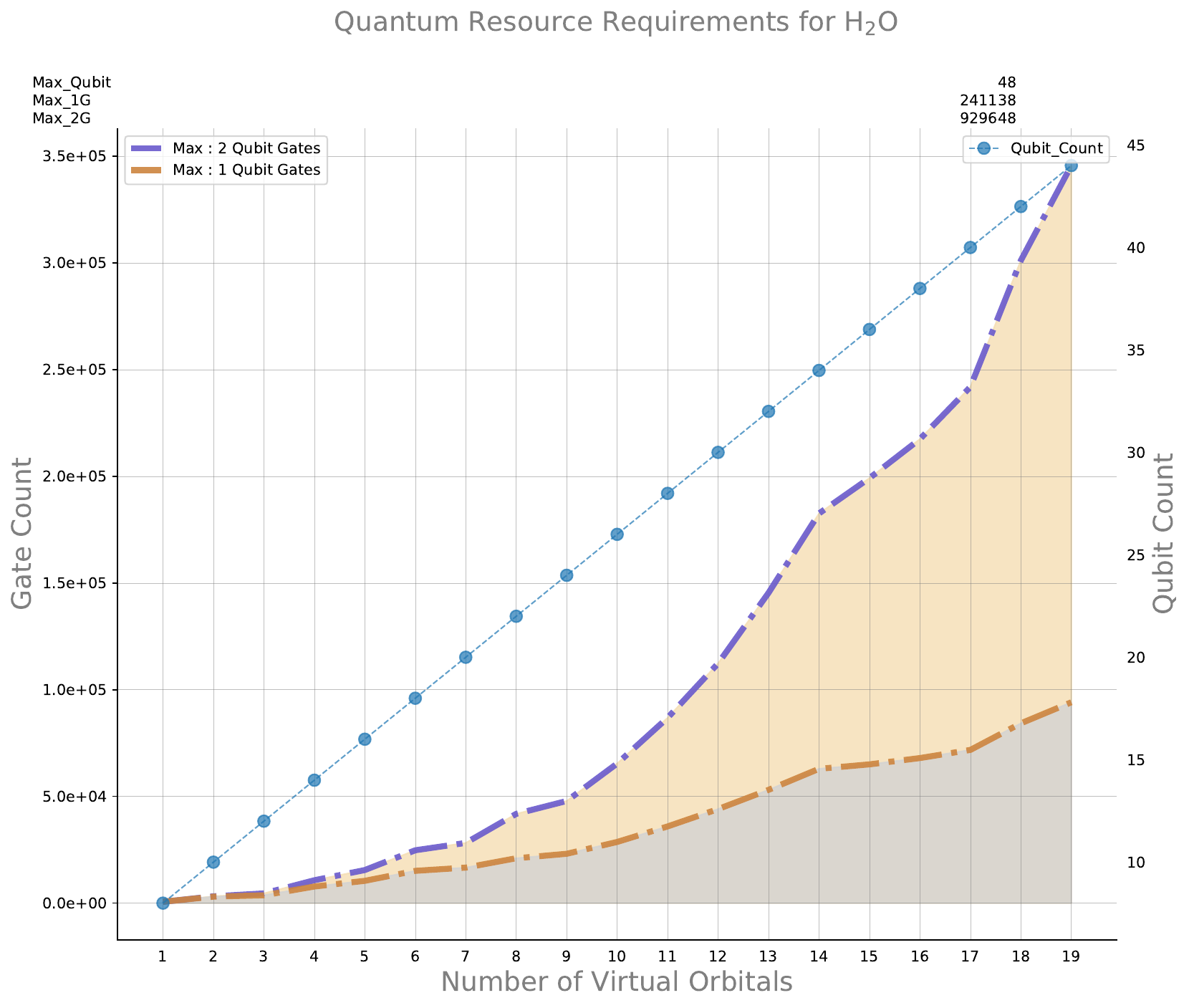}
	\caption{Quantum resources required to prepare the quantum state of the subsystem of H$_2$O as a function of the number of virtual orbitals}
	\label{fig:Quantum_Resources_H2O}
\end{figure*}

\begin{figure*}[tbp]
	\includegraphics[scale=0.38]{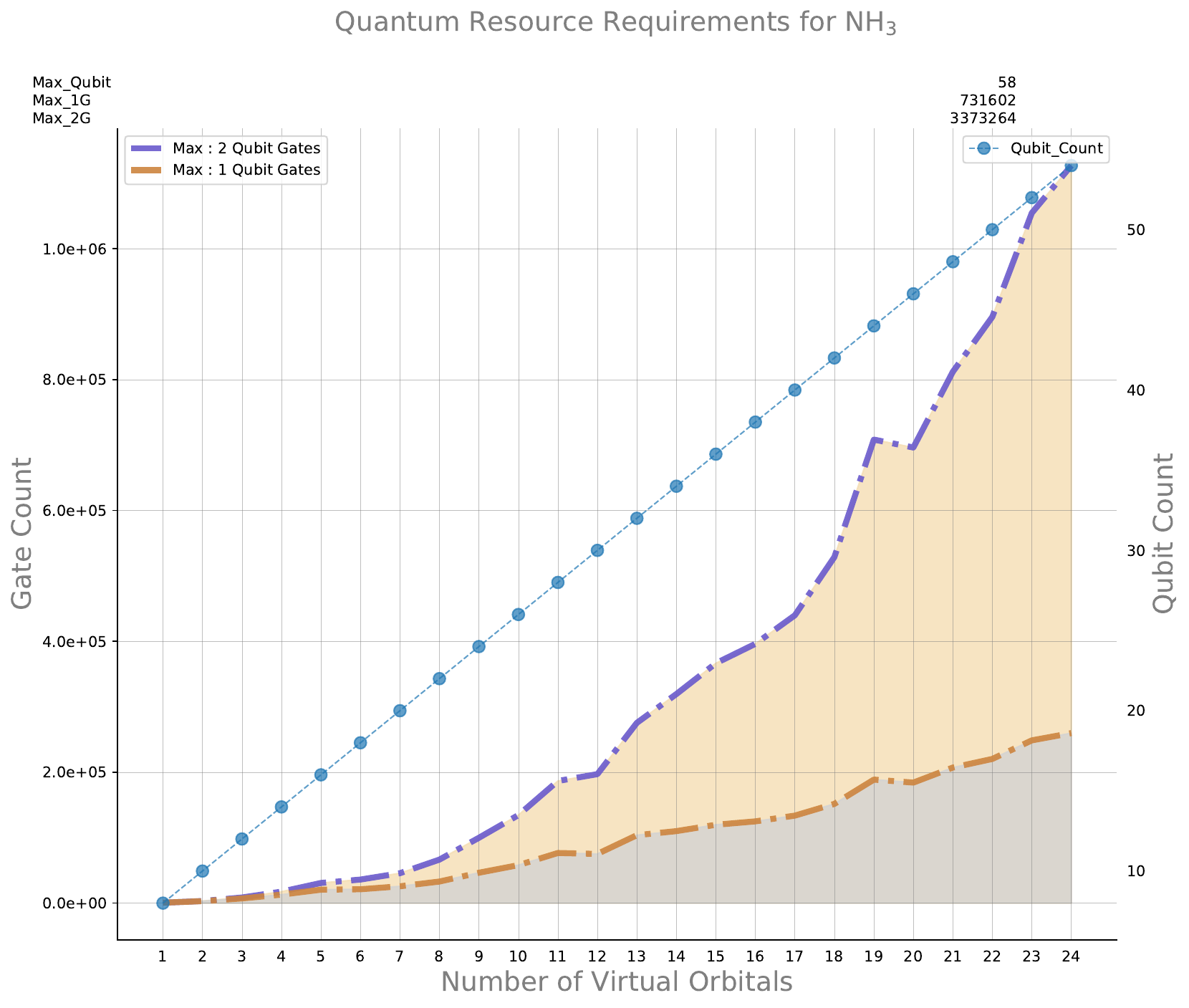}
	\caption{Quantum resources required to prepare the quantum state of the subsystem of NH$_3$ as a function of the number of virtual orbitals}
	\label{fig:Quantum_Resources_NH3}
\end{figure*}

\begin{figure*}[tbp]
	\includegraphics[scale=0.38]{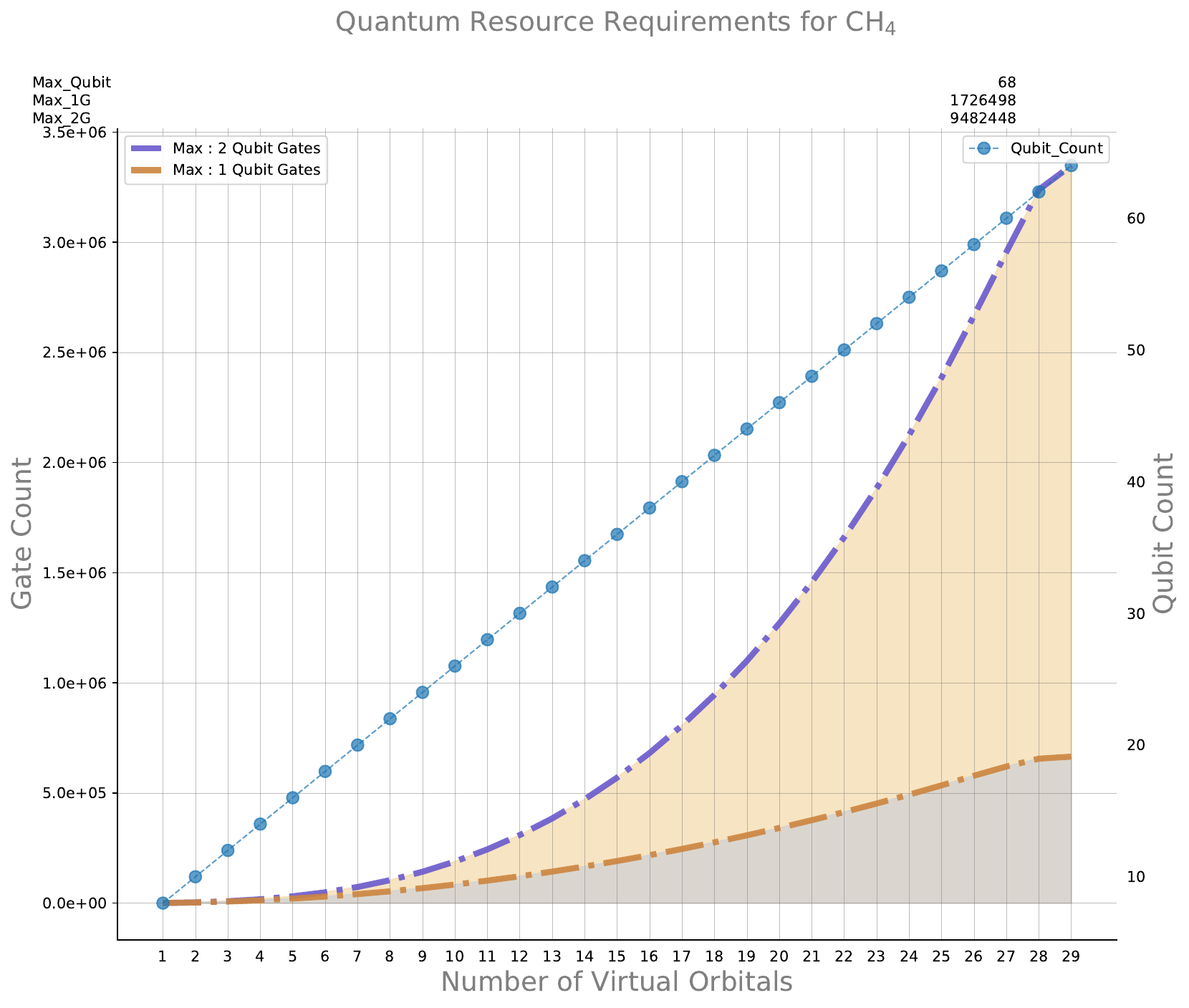}
	\caption{Quantum resources required to prepare the quantum state of the subsystem of CH$_4$ as a function of the number of virtual orbitals}
	\label{fig:Quantum_Resources_CH4}
\end{figure*}

Figures~\ref{fig:Quantum_Resources_BeH2} to~\ref{fig:Quantum_Resources_CH4} show the quantum resources required to prepare the quantum states of the subsystems for each of the molecules BeH$_2$, HF, H$_2$O, NH$_3$, and CH$_4$ as a function of the number of virtual orbitals. The MI(\(n\))-FNO approach (where $n$ is restricted up to three-body terms) produces many increments, and the quantum resources that are required for each vary depending on the increment size. The largest numbers of qubits and one- and two-qubit gate counts are plotted for the increment that has the largest quantum resource requirements.

\clearpage

\section{{Effectiveness of FNO in Compacting Virtual Space of a CGC catalyst}} \label{appendix:appC}
\begin{figure*}[h]
	\includegraphics[scale=0.44]{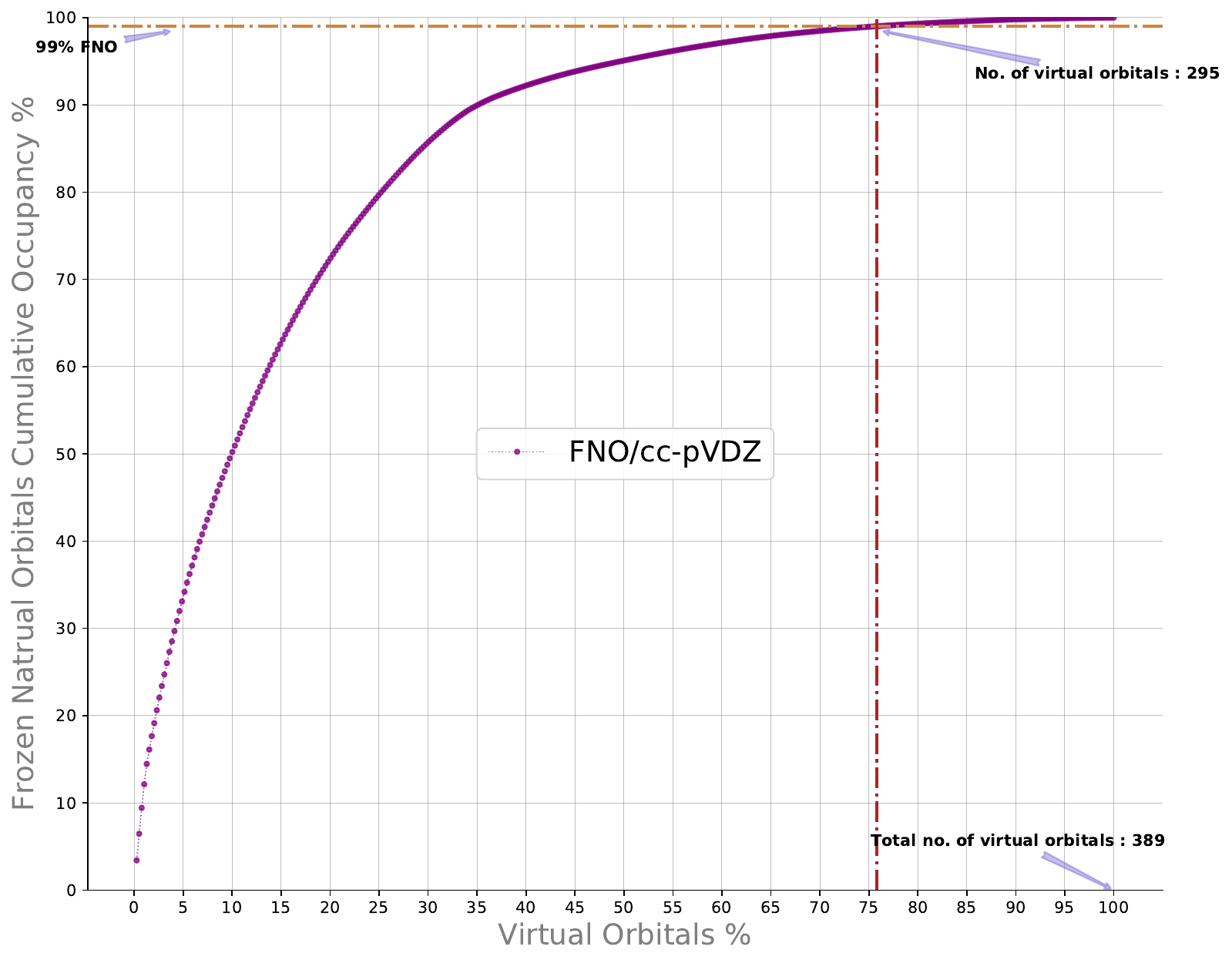}
	\caption{Cumulative FNO occupancy as a function of the number of virtual orbitals for a CGC catalyst for FNO-CCSD/cc-pVDZ}
	\label{fig:cgc_FNO_DZ}
\end{figure*}

\begin{figure*}[h]
	\includegraphics[scale=0.44]{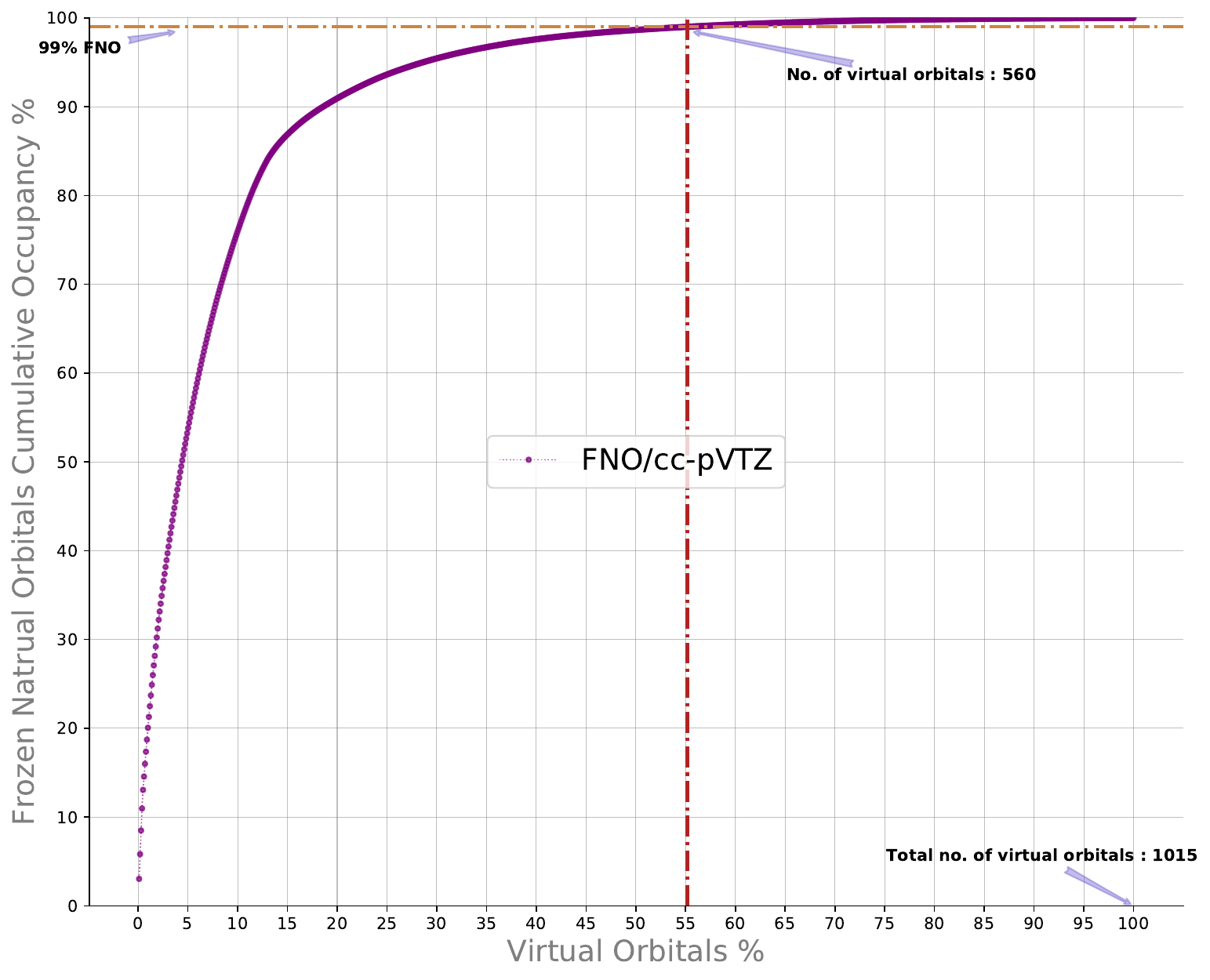}
	\caption{Cumulative FNO occupancy as a function of the number of virtual orbitals for a CGC catalyst for FNO-CCSD/cc-pVTZ}
	\label{fig:cgc_FNO_TZ}
\end{figure*}

\begin{figure*}[h]
	\includegraphics[scale=0.44]{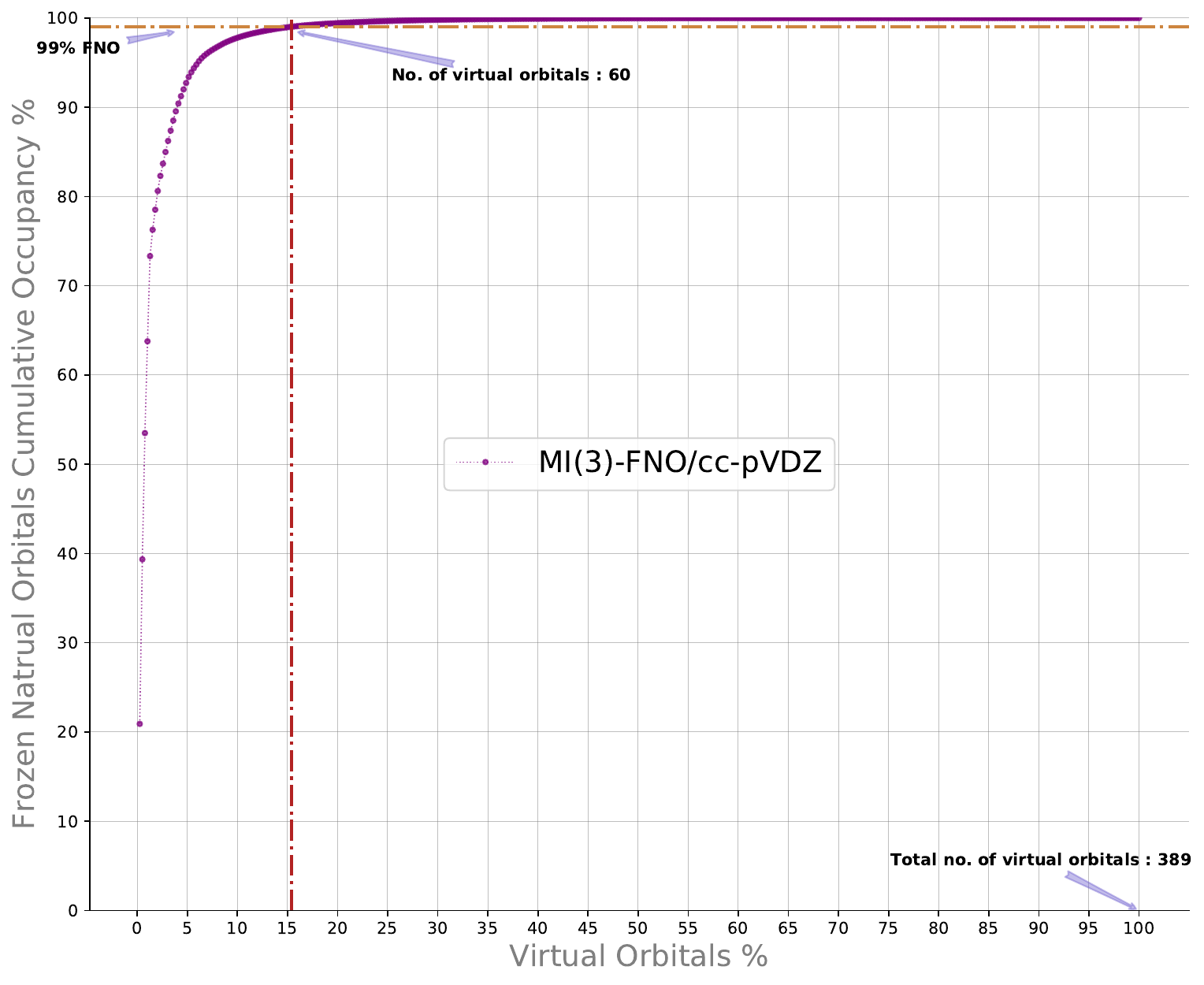}
	\caption{Cumulative FNO occupancy as a function of the number of virtual orbitals for a CGC catalyst for MI(3)-FNO-CCSD/cc-pVDZ}
	\label{fig:cgc_MI_DZ}
\end{figure*}

\begin{figure*}[h]
	\includegraphics[scale=0.44]{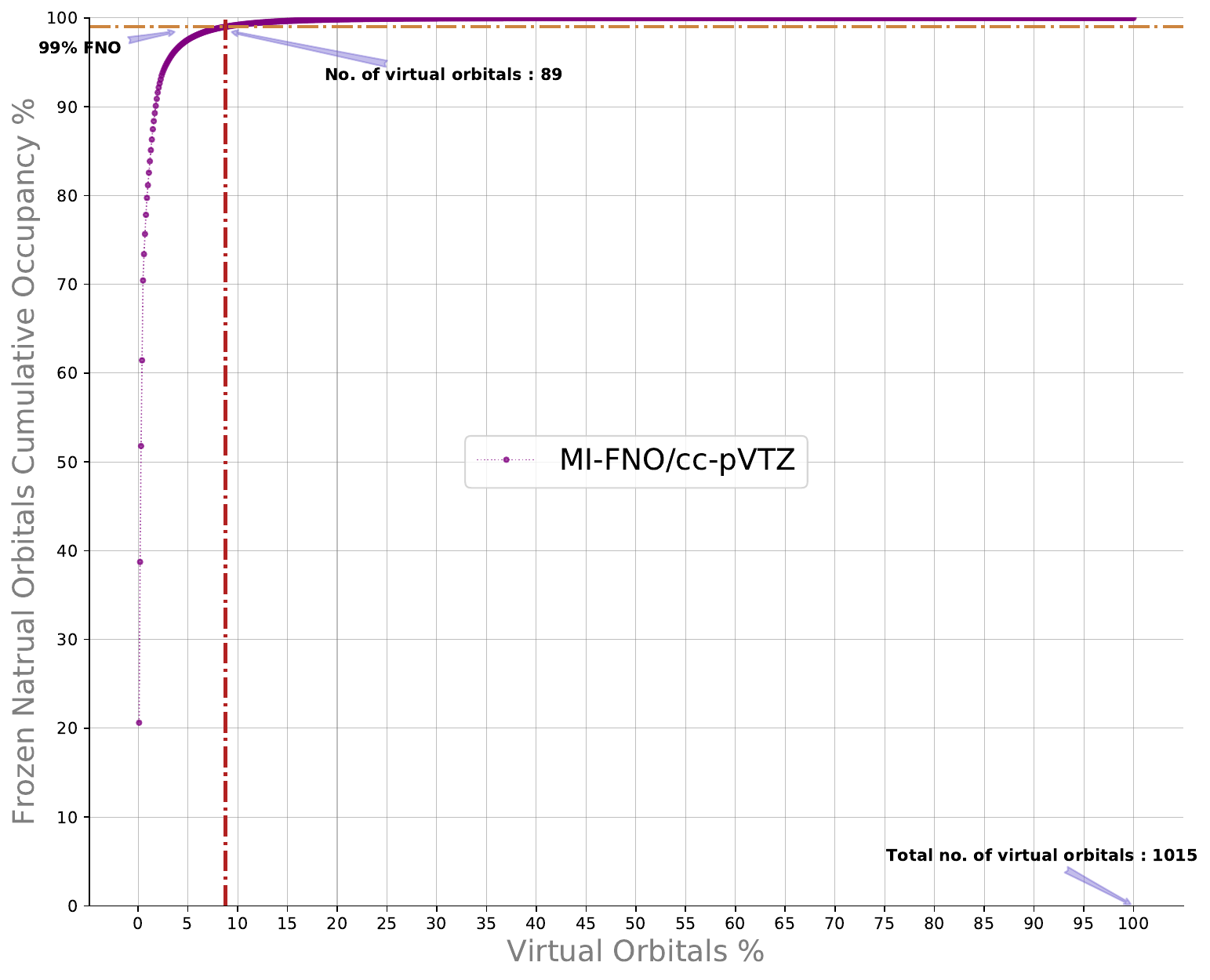}
	\caption{Cumulative FNO occupancy as a function of the number of virtual orbitals for a CGC catalyst for MI(3)-FNO-CCSD/cc-pVTZ}
	\label{fig:cgc_MI_TZ}
\end{figure*}

Figures~\ref{fig:cgc_FNO_DZ}, \ref{fig:cgc_FNO_TZ}, \ref{fig:cgc_MI_DZ}, and \ref{fig:cgc_MI_TZ} show the cumulative FNO occupancy percentage as a function of the number of virtual orbitals for a CGC catalyst. The values on the horizontal axis represent the ratio of the number of virtual orbitals, calculated as ``the number of virtual orbitals that are used in the calculation'' divided by ``the total number of virtual orbitals of the system'', while the vertical axis represents the cumulative FNO occupancy percentage. The dash-dotted horizontal line shows the FNO occupancy of 99\%. The plots are obtained by running  FNO or MI-FNO calculations with either a cc-pVDZ or cc-pVTZ basis.

\end{document}